\documentstyle{article}
\input epsf
\begin{document}
\rightline{EFI-95-62}
\rightline{hep-th/9509149}
\vskip 1cm
\centerline{\LARGE \bf String field theory in curved spacetime}
\centerline{\LARGE \bf and the resolution of spacelike singularities}
\vskip 1cm

\centerline{{\bf \Large Albion Lawrence and Emil Martinec}}
\vskip .5cm
\centerline{\it Enrico Fermi Inst. and Dept. of Physics}
\centerline{\it University of Chicago,
5640 S. Ellis Ave., Chicago, IL 60637 USA}

\begin{abstract}

We attempt to understand the fate of spacelike gravitational singularities
in string theory via the quantum stress tensor
for string matter in a fixed background.  We first approximate
the singularity with a homogeneous
anisotropic background and review the minisuperspace
equations describing the evolution of the
scale factors and the dilaton.  We then review and discuss the
behavior of large strings in such models.  In a simple model which
expands isotropically for a finite period of time we compute the
number density of strings produced by quantum pair production
and find that this number, and thus the stress tensor,
becomes infinite when the Hubble volume of the expansion exceeds
the string scale, in a manner reminiscent of the Hagedorn transition.
Based on this calculation we argue that either the region near
the singularity undergoes a phase transition when the density
reaches the order of a string mass per string volume, or that the
backreaction of the produced string matter dramatically modifies
the geometry.

\end{abstract}

\def\figloc#1#2{\vbox{{\epsfxsize=4.5in
        \nopagebreak[3]
    \centerline{\epsfbox{fig#1.ps}}
        \nopagebreak[3]
    \centerline{Figure #1}
        \nopagebreak[3]
    {\raggedright\it \vbox{  #2 }}}}
    \bigskip
    }
\def\pref#1{(\ref{#1})}

\def\ie{{\it i.e.}}
\def\eg{{\it e.g.}}
\def\cf{{\it c.f.}}
\def\etal{{\it et.al.}}
\def\etc{{\it etc.}}

\def\adv{{\it Adv. Phys.}}
\def\ap{{\it Ann. Phys, NY}}
\def\cqg{{\it Class. Quant. Grav.}}
\def\cmp{{\it Commun. Math. Phys.}}
\def\jetp{{\it Sov. Phys. JETP}}
\def\jetpl{{\it JETP Lett.}}
\def\jp{{\it J. Phys.}}
\def\ijmp{{\it Int. J. Mod. Phys. }}
\def\nc{{\it Nuovo Cimento}}
\def\np{{\it Nucl. Phys.}}
\def\mpl{{\it Mod. Phys. Lett.}}
\def\pl{{\it Phys. Lett.}}
\def\pr{{\it Phys. Rev.}}
\def\prl{{\it Phys. Rev. Lett.}}
\def\prcl{{\it Proc. Roy. Soc.} (London)}
\def\rmp{{\it Rev. Mod. Phys.}}

\def\p{\partial}

\def\apr{\alpha'}
\def\str{{str}}
\def\lstr{\ell_\str}
\def\gstr{g_\str}
\def\Mstr{M_\str}
\def\lpl{\ell_{pl}}
\def\Mpl{M_{pl}}
\def\varep{\varepsilon}

\def\AA{{\cal A}}
\def\BB{{\cal B}}
\def\CC{{\cal C}}
\def\DD{{\cal D}}
\def\EE{{\cal E}}
\def\FF{{\cal F}}
\def\GG{{\cal G}}
\def\HH{{\cal H}}
\def\II{{\cal I}}
\def\JJ{{\cal J}}
\def\KK{{\cal K}}
\def\LL{{\cal L}}
\def\MM{{\cal M}}
\def\NN{{\cal N}}
\def\OO{{\cal O}}
\def\PP{{\cal P}}
\def\QQ{{\cal Q}}
\def\RR{{\cal R}}
\def\SS{{\cal S}}
\def\TT{{\cal T}}
\def\UU{{\cal U}}
\def\VV{{\cal V}}
\def\WW{{\cal W}}
\def\XX{{\cal X}}
\def\YY{{\cal Y}}
\def\ZZ{{\cal Z}}

\section{Introduction}

Infinities and singularities in physical theories are
signals that the theory
in question is inconsistent or incomplete, and force physicists to
uncover deeper structures.  Much of twentieth century physics has
arisen from such singularities; two notable examples are
the ultraviolet catastrophes of classical
thermodynamics which helped drive the development of
quantum mechanics, and the infinities in field theory
loop calculations
which helped drive the development
and understanding of the renormalization group.

Among the theories which we currently use to
describe the world,
general relativity remains doubly afflicted:
at the quantum level it
is nonrenormalizable, and
at the classical level it contains
singularities, as demonstrated by the well-known singularity theorems
of Hawking and Penrose (Penrose 1965, Hawking 1967,
Hawking and Penrose 1970).  With both of these singularities we are
faced with the problem of predictability.  Nonrenormalizability
tells us that there
must be a very different description of the world
at the Planck scale, but gives no clue
as to what that description might be.  Classical spacetime singularities
defined via geodesic incompleteness
tell us that in finite proper time, we will reach regions of spacetime
beyond which we cannot evolve the dynamical equations of the
theory.  Of course,
if we attempt to cure this classical problem by quantizing the theory,
we must contend with the nonrenormalizability of the theory.

One subject of active research which may force physicists to face these
two problems directly is the apparent evaporation of black
holes via quantum particle production
(Hawking 1974).  It is our belief that to understand the
outcome of the Hawking process we must eventually understand
the physics near the singularity.  In particular this seems to
be the primary lesson taught by the study of two-dimensional
dilatonic black holes; at some point the singularity and
the apparent horizon coincide and one needs some prescription
for dealing with the singularity at this point in order to understand
the final state of the black hole (see Strominger 1995 and references therein).
To date there have been several proposals
for the what endpoint the of black hole evaporation process
might be\footnote{``The
number of very good physicists who have expressed
fairly definitive opinions about the resolution of the Hawking
puzzle is smaller than the number of definitive opinions they
have expressed.'' (Banks 1994)}\ (for
critical reviews and references see Banks 1994 and Strominger 1995);
many of these proposals rely on low-energy physics up to the point
that the black hole reaches the Planck mass, and then propose
plausible consistent completions of the evaporation process,
without recourse to a Planck-scale
theory.  Nonetheless, it remains to be shown explicitly that any
of these proposals are
the result of a complete theory of quantum gravity.

Thus, if we claim to understand quantum gravity, we must understand and
remove the singularities that seem to plague it.  This is hardly
a new enterprise, and while we are not capable of producing a full
history of previous attempts to do so, one may divide
the attempts into two categories:
\begin{itemize}
\item{Evasion of some tenet of the singularity theorems}
\item{Searching for a breakdown of the notion of classical
spacetime in a region of strong curvature, via quantum mechanics
or string theory}
\end{itemize}
The most natural way to avoid the singularity theorems
is to violate the condition
\begin{equation}
\RR_{a b}\xi^{a}\xi^{b} \geq 0,
\label{posit}
\end{equation}
(where $\RR_{ab}$ is the Ricci tensor and $\xi$ is a timelike or null
vector field depending on which theorem one appeals to); the
other conditions involve global hyperbolicity of the spacetime
(although this condition is non-essential), the lack of closed timelike loops,
and the existence of trapped surfaces (which occurs given a sufficient
concentration of matter (Schoen and Yau 1983)), and are physically too
attractive to casually dismiss.  The singularity theorems
are based on feeding Equation \pref{posit} back into the Raychaudhuri
equation; in general relativity
one can then use Einstein's equations to relate Equation \pref{posit} to
the strong or the null energy conditions.  To avoid the
singularity theorems one must then either
violate the appropriate energy conditions, or suitably modify
Einstein gravity so that Equation \pref{posit} is violated.
In particular, quantum fields violate the energy conditions
(Epstein \etal\ 1965).

In the long run simply adding terms to Einstein's
equations or relaxing energy conditions may not seem entirely satisfactory:
we know that as we approach large field gradients
close to the singularity,
quantum geometry and (sub)Planckian physics come into play.
Still, at this point
the burden is on the theorist to
explain what exactly quantum geometry is, and then to show
that quantum mechanics combined with
one's favorite choice of a Planck scale theory is not also singular,
especially given the nonrenormalizability
of Einstein gravity.

In this paper we would like to examine gravitational singularities
using quantum string theory.  To do this we should redefine
singularities in terms of extended
objects and quantum mechanics; we can
do this with fairly natural extensions of
geodesic completeness.  Classically, we require that the
solutions to the string equations of motion never cause the string to
leave the spacetime in finite proper worldsheet distance.
Quantum mechanically, we require that the
first-quantized wave operator be well-defined and Hermitian
everywhere.
With these definitions we can discuss three types of singularities:
singular sources of classical fields (\ie\ charged point particles),
timelike singularities, and spacelike singularities
occurring (in a suitably defined coordinate system) at some point in time.

Singular classical sources are easily dealt with:
they arise generally from considering particles as
point-like objects.  In a first-quantized theory quantum
mechanics smears out the wave function of a test particle enough to avoid
any pathology at the source, so long as the singularity in the
potential is weaker than $1/r^{2}$ (which is
the case for the known forces of nature.)  In quantum field theory
calculations, we get divergences from large loop momenta
and from small impact parameters.  These are results
of not properly regulating the theory at short distances.

Timelike singularities, of which the aforementioned singularities
are a subset, are also not generically troublesome.
It is strongly believed that string theory has a minimum distance and
thus soft high-energy behavior, due to the extended
nature of the fundamental objects.  This has been shown
to resolve many potential timelike singularities in the theory, most notably
in amplitudes for high-energy scattering (Gross and Mende 1987, 1988),
orbifold singularities (Dixon \etal\ 1985, 1986), and singularities
arising from the degeneration of K\"{a}hler structures
in Calabi-Yau manifolds (Aspinwall \etal\ 1993 and 1994, Witten 1993).
This good behavior arises from
the apparent lack of any operative definition
of a point in string theory: no point on a given string can
carry a finite fraction of the string's energy, so that any momentum
transfer in a stringy scattering process is shared by the entire string.
We should note that we do not want to smooth
out {\it all} timelike singularities. Horowitz and Myers (1995) have
argued that any stable theory with Einstein gravity as a
low-energy limit must have solutions with
timelike singularities arising from singular initial conditions;
in particular, a solution which looks like a negative-mass
Schwarzschild solution at large distances
should be singular even in a full theory of
quantum gravity (remember that the total mass
is measured at infinity, where only the Einstein term
in the effective action will contribute).  If
we cannot throw away such solutions the theory becomes unstable since
there is no stable ground state, \ie\ empty space decays by pair-producing
these negative-energy lumps (Horowitz and Myers, 1995).

Spacelike singularities
pose a genuinely serious and confusing dilemma,
as they occur entirely because
of the universally attractive nature of gravity
past the event horizon, where the causal structure
forces classical matter to collapse completely.
These facts seem to be independent
of the ultraviolet nature of the theory.  So, while the singularity
is an ultraviolet object and perhaps will be resolved by physics
similar to that which resolves timelike singularities, their inevitable
nature and the fact that they arise from a rather soft attractive
force begs for a careful examination.  At any rate it is safe to say
that we do not have as much
control over these singularities in string theory
as we have over the known timelike singularities,
and so those of us who hope that string theory describes nature
must still confront spacelike singularities.

The first words uttered by string theorists when faced with
singular backgrounds is that they are simply signs that the
expansion of the spacetime effective action in
powers of $\ell_{{\rm curv.}}/\lstr$,
where $\ell_{{\rm curv.}}$ is the characteristic length over which
the curvature changes, breaks down.  Yet there are known {\em exact}\
conformal theories that possess spacelike singularities, the
best known example being the 2-dimensional $SL(2,\RR)/U(1)$ black hole,
(Witten 1991).
One might then appeal to string loop corrections to the
$\beta$-functions
(Lovelace 1986, Fischler and Susskind 1986, Callan \etal\ 1987).
While we will argue below that string loop effects are part of the story,
we should
be careful.  For example, there are gravitational plane-wave backgrounds with
null singularities, for which string propagation is singular
(Horowitz and Steif 1990a and 1990b, de Vega and Sanchez 1992). These
backgrounds correspond to
exact conformal theories at tree level (Amati and Klimcik 1989, Horowitz and
Steif 1990a) because all interesting counterterms that one might
produce from the Riemann tensor, the dilaton, or the antisymmetric
tensor field, will vanish; for the same reason loop
corrections to the background will also vanish.  However, it should
be noted that these null singularities by definition do not arise from
nonsingular initial conditions.
We should also recall the arguments of Horowitz and Myers (1985)
mentioned above: if corrections to the low energy effective
action generically
tame singular backgrounds, we should worry that the theory
is unstable.

The aforementioned string theorist would next point out that
corrections to the effective action
are not the end of the story; for example, strings propagating in orbifold
geometries are quantum mechanically well behaved at the orbifold
points.  The lesson here is that we should relax our definition
of singularities somewhat, and define them as regions of the
background for which the wave operator for string
theory becomes undefined, thus properly including orbifold
geometries and some degenerations of K\"{a}hler geometries
as nonsingular backgrounds.  This definition is a natural
extension of geodesic completeness as it defines a singularity as
the point at which predictability, here defined as the ability to evolve
quantum wave functions, breaks down.
This relaxed definition does not save us from spacelike
singularities.  For example, the $L_{0}$ operator
in the $SL(2,\RR)/U(1)$ black hole background becomes ill-defined at
the singularity: string
wave functions (\ie\ tachyon vertex operators) blow
up at the singularity (Dijkgraaf \etal\ 1992).  In order
to support our claim that $\apr$ corrections and the properties
of quantum wave functions in string theory do not
in general save us from spacelike singularities,  we will
in the appendix review arguments which indicate that the
structure of higher-dimensional black hole singularities
is described by a perturbation of the $SL(2,\RR)/U(1)$ black hole.
Note also that the caveats of Horowitz and Myers (1995) apply
to quantum mechanical smoothing as well.

None of this is to say that there is no solution to the problem of
spacelike singularities in string theory, or that the answer does not
involve the short-distance properties of string theory; rather,
we simply wish to emphasize that
the problem has not been solved and that it requires some care.
In the light of these arguments,
this paper is an attempt to reach a primitive understanding of the quantum
dynamics of string field theory in singular backgrounds, as a
precursor to understanding the full quantum evolution of the background.
We will make this attempt
by first examining the spacetime background near the singularity and noting
its generic and essential features, and then by examining
string propagation in singular backgrounds, and
finally by trying to crudely understand string field theory in a simplified
spacetime background for which the scale factor changes over a brief
period of time.  We will find that the density of produced strings
in a given state is
exponentially suppressed by the mass of that state
times the inverse rate
of expansion, $\rho^{-1}$, so that when this rate becomes as large
as $\lstr^{-1}$, the density of states overwhelms the
suppression and an infinite number of highly energetic long strings are
produced.  One may object that it is not legitimate
to quantize string field theory in a classical background
spacetime without explicitly taking backreaction
into account from the start; however, since string theory
has an as adjustable dimensionless parameter the ratio $\lstr/\lpl$
far from the black hole, we may make this parameter small and see if
string theory resolves the singularity in the
weakly coupled region where we have some hope of knowing what to do.
Our result indicates a sort of paradox, of which we see two
possible resolutions: either stringy
backreaction must smooth the singularity, or there is a Hagedorn-like
phase transition as
the density of produced strings becomes comparable to a
string mass per unit string volume.
In the former case we will suggest that the combination of
an equation of state for string matter derived by
Gasperini \etal\ (1991a,b) and the effects of the dilaton
(Tseytlin and Vafa 1992) potentially cause spacetime to become non-singular.
Without thinking too much, one could imagine a few scenarios.
For example, the spacetime
could stabilize as expansion or contraction cease, with
the dilaton kinetic energy balancing the produced energy density as
the spacetime runs to strong coupling.  Another
possibility is that the singular region relaxes to
some de-Sitter like phase, as
suggested by Frolov \etal\ (1990) (see also
Poisson and Israel 1988) for the case of particle field theory,
based on a limiting-curvature hypothesis for
quantum gravity.  One of us (Martinec 1995) proposed a
realization of this latter possibility by appealing to T-duality:
this paper arose from that work, although we no longer
feel that the duality properties of string theory are
necessarily relevant
for our purposes.
Our arguments will be crude and will
ignore many issues such as the conformal invariance of the
background, but we will argue that we have captured the physics of
the effects we are interested in and that what we have ignored
will not invalidate our results.

Since our proposals depend on the time-dependent nature of the
singularities we are interested in, they may not apply to
timelike singularities; the sorts of timelike
singularities which Horowitz and Myers (1995) discuss
may remain singular as required for the stability of flat space.
As a limiting case,
the singular null-fronted plane wave spacetimes described by
Horowitz and Steif (1990) and de Vega and Sanchez (1992)
will have no particle creation (and no string
creation) due to the presence of a null Killing vector, so
that in this borderline case the singularity is untamed by
our scenarios.

The remainder of the paper shall be organized as follows:
in Section 2 we will review features of singular backgrounds
in general relativity
and string theory.  In the case of string theory, we will
examine the cosmological equations of Tseytlin and Vafa (1992)
with respect to various phenomenological energy-momentum tensors that one may
imagine, and discuss the singularities that may arise in these equations.
In Section 3 we will describe string propagation
in expanding and/or contracting backgrounds, reviewing
the work of Gasperini,
Sanchez, and Veneziano (Sanchez and Veneziano 1990, Gasperini \etal\ 1991a,b).
We will use a similar approximation to examine the
propagation of large strings in a temporarily expanding model.
We will also discuss the
equation of state for string matter in these models.  In Section 4
we will discuss free quantum string field theory in curved backgrounds,
and describe techniques for computing approximations to
the rate of string production in such backgrounds.
In Section 5 we will apply these techniques to a specific
isotropically expanding model and find the aforementioned singularity
in the rate of string production.  In section 6 we will discuss how one
might extend this calculation to an anisotropic model and what
features one might look for; we will also discuss how the produced
matter, given the equation of state discussed in section 3, might
affect the singularity.  Section 7 will contain our conclusions and
further speculations.  Appendix A will discuss how one might
relate the $SL(2,\RR)/U(1)$ black hole to the Schwarzschild solution.

\section{Classical backgrounds in general relativity and string theory}

Belinskii, Khalatnikov, and Lifshitz (Lifshitz and Khalatnikov 1963,
Belinskii \etal\ 1970, 1982) have argued that the spacetime metric
close to a spacelike singularity
can be written as
\begin{equation}
ds^{2} = -dt^{2} + \left(t^{2p_{1}}l_{\alpha}l_{\beta}
+ t^{2p_{2}}m_{\alpha}m_{\beta} + t^{2p_{3}}n_{\alpha}n_{\beta}
\right) dx^{\alpha}dx^{\beta}\ ,
\label{singmet}
\end{equation}
where $p_{i}$, $l$, $m$, and $n$ are functions of the spatial
coordinates only, and the $p_{i}$ satisfy the equations
\begin{equation}
\sum_{i=1}^{3} p_{i} = \sum_{i=1}^{3} p_{i}^{2} = 1\ ,
\label{kasexp}
\end{equation}
which are identical to the restrictions on the spatially independent
exponents $p_{i}$ in
a vacuum Kasner spacetime:
\begin{equation}
ds^{2} = -dt^{2} + \sum_{i=1}^{3} t^{2p_{i}}\left(dx^{i}\right)^{2}\ .
\end{equation}
Equation \pref{singmet} is found by expanding Einstein's equations
in a power series in $t$, assuming a power law singularity in the metric,
with the leading exponents as shown: one finds that
contributions from the intrinsic curvature of the spatial slices
are subleading as compared to their extrinsic curvature
(though there is the requirement that, if $p_{1}$
is the negative exponent, $\vec{l}\cdot\nabla\times\vec{l}=0$,
which means that we may use the direction specified by $l$ as
a global coordinate).  This
means that the spacetime is ``velocity-dominated''
in this region.  If we look on a reasonably small spatial
slice, the metric should look like the Kasner metric.
The form \pref{singmet} is certainly what we find near the Schwarzschild
singularity: the metric inside the horizon ($r<r_{h}$) for
a D-dimensional Schwarzschild black hole is
\begin{eqnarray}
ds^{2}&=&-\left[\left(\frac{r_{h}}{r}\right)^{D-3} - 1\right]^{-1}dr^{2}
+\left[\left(\frac{r_{h}}{r}\right)^{D-3}-1\right]
dt^{2}+r^{2}d\Omega_{D-2}^{2}
\nonumber\\
&\sim& -\left(\frac{r}{r_{h}}\right)^{D-3}dr^{2}
+\left(\frac{r_{h}}{r}\right)^{D-3}dt^{2} + r^{2}d\Omega_{D-2}^{2}\ .
\end{eqnarray}
If we write $T = -[2r_{h}/(D-1)](r/r_{h})^{(D-1)/2}$, and $\rho = t$,
we can rewrite the most singular part of the metric as
\begin{equation}
ds^{2} = - dT^{2} + \left[\frac{(D-1)T}{2r_{h}}\right]^{-2(D-3)/(D-1)}
d\rho^{2} + r_{h}^{2}\left[\frac{(D-1)T}{2r_{h}}\right]^{4/(D-1)}
d\Omega_{D-2}^{2}\ .
\end{equation}
Note that in four dimensions the relations \pref{kasexp} are satisfied.
We conclude that the essential physics of the
velocity-dominated
singularity can be captured by looking at homogeneous anisotropic
spacetimes.

Einstein's equations are contained in the lowest order
$\beta$-function equations of string theory (Lovelace 1986,
Callan \etal\ 1985,
Fradkin and Tseytlin 1985, Sen 1985);
one does not expect the Schwarzschild solution to describe
the geometry as one approaches the singularity, where higher order
terms in the $\beta$-function equations become important.  Nonetheless it is
worth repeating that there are exact conformal theories which
contain spacelike singularities, including the $SL(2,\RR)/U(1)$
black hole (Witten 1991), various four dimensional
black hole backgrounds consisting of the
tensor product of this theory with other
conformal field theories (Giddings \etal\ 1993, 1994, Johnson 1994,
Lowe and Strominger 1994),
and a four dimensional cosmological model, with initial and final
singularities,
constructed as a c=4 WZW model (Nappi and Witten 1992).
Furthermore, there are some indications that the $SL(2,\RR)/U(1)$
black hole
may be relevant for understanding the background near the singularity
in higher-dimensional spacetimes; we will review these indications
in the appendix.

We will start with something rather simpler than these exact conformal
field theories, looking instead for
homogeneous solutions to the one-loop beta functions for the metric
coupled to the dilaton $\phi$ and matter with
central charge
\begin{equation}
c=\frac{2}{3}\left(d_{{\rm crit}} - D - N_{{\rm compact.}}\right)\ ,
\end{equation}
where $d_{{\rm crit}}$ is the critical dimension of the string theory,
$D$ is the number of macroscopic dimensions, and
$N_{{\rm compact.}}$ is the number of compactified dimensions.
If we write the string frame metric as
\begin{equation}
ds^{2} = -dt^{2} + \sum_{i}e^{2\lambda_{i}(t)} \left(dx^{i}\right)^{2}\ ,
\end{equation}
then the $\beta$-function equations can be written as (Tseytlin and Vafa 1992)
\begin{eqnarray}
&&c-\sum_{i=1}^{D-1}\left(\dot{\lambda}_{i}\right)^{2}+
\left(2\dot{\phi}-\sum_{i=1}^{D-1}\dot{\lambda}_{i}\right)^{2}
= \rho e^{a\phi} \nonumber \\
&&\ddot{\lambda}_{k} -
\left(2\dot{\phi}-\sum_{i=1}^{D-1}\dot{\lambda}_{i}\right)\dot{\lambda}_{k}
= e^{a\phi}\frac{p_{k}}{2} \nonumber \\
&&2\ddot{\phi} - \sum_{i=1}^{D-1}\ddot{\lambda}_{i}
-\sum_{i=1}^{D-1}\dot{\lambda}_{i}^{2} = e^{a\phi}\frac{\rho}{2}\ .
\label{stcoseq}
\end{eqnarray}
Here $\rho$ is the energy density, and $p_{k}$ is the pressure
in the direction of $x^{k}$: these terms come from adding a phenomenological
matter action to the low energy effective action that can be derived
from the usual $\beta$-function equations.  We may also include
the effect of potential terms, such as that coming from a
dilaton potential, in $\rho$ and $p$.
In general $a$ will depend on how the matter couples to the
dilaton.  For a classical fundamental string source, the total action
without the antisymmetric tensor field is
(Dabholkar and Harvey 1989, Dabholkar \etal\ 1990)
\begin{eqnarray}
\lefteqn{S = \frac{1}{16\pi G} \int d^{D}x \sqrt{g} e^{-2\phi}
\left\{\RR + 4 \left( \nabla \phi \right)^{2}\right\}}     \nonumber\\
&&- \frac{1}{2\pi\apr}\int d^{2}\sigma\left\{\sqrt{h}h^{\alpha\beta}
g_{\mu\nu}\p_{\alpha}X^{\mu}\p_{\beta}X^{\nu} + \frac{\apr}{2}
\RR^{(2)}\phi(X)\right\} \nonumber \\
&&= S_{{\rm spacetime}} + S_{{\rm worldsheet}}[X]\ ,
\label{dabhact}
\end{eqnarray}
where $\RR$ is the spacetime curvature, $\RR^{(2)}$
is the worldsheet curvature, and $X(\sigma, \tau)$ is the location of
the worldsheet.  Here the fundamental string couples as a classical
source, and thus lacks the $\exp (-2\phi) = 1/\gstr^{2}$ factor.
(We will discuss the justification for
Equation \pref{dabhact} at the end of the following section.)  Naturally, for
multiple strings, one may sum the worldsheet action over the different
sources, making the replacement
\begin{equation}
S_{{\rm worldsheet}}[X] \longrightarrow \sum_{i} S_{{\rm worldsheet}}[X_{i}]
\end{equation}
in Equation \pref{dabhact}.
Now if the stress tensor arises from a gas of classical strings,
we expect that $a=2$ due to the lack of a dilaton prefactor in the
worldsheet action (Veneziano 1991, Tseytlin 1992).
We will also get different values of $a$ depending on whether the matter
comes from higher loop effects and so on
(Tseytlin 1992).  In the particle production scenario we are proposing,
the number density will be proportional to $\exp(-\mbox{const.}
/\dot{\lambda})$,
so that we will set $a=2$ and let $\rho = \rho(\phi,\lambda,\dot{\lambda})$
and $p = p(\phi,\lambda,\dot{\lambda})$.
It is worth noting that if we set the dilaton to be constant
then Equation \pref{stcoseq} has solutions only if
\begin{equation}
\rho=\sum_{i=1}^{D-1}p_{k} + 2c\ .
\end{equation}
In particular, for $c=0$ a constant dilaton and a homogeneous
metric requires a radiation-like equation of state.

Many solutions to Equation \pref{stcoseq} have been described
by Tseytlin (1992): we would like to prepare for later discussions
by sketching a few possible solutions.  Define a new field
$\varphi=2\phi-\sum_{i=1}^{n}\lambda_{i}$ ,
and set $a=2$; then Equation \pref{stcoseq} becomes
\begin{eqnarray}
&&c-\sum_{i=1}^{n}\dot{\lambda}_{i}^{2}+\dot{\varphi}^{2}=
e^{2\phi}\rho\nonumber\\
&&\ddot{\lambda}_{k}-\dot{\varphi}\dot{\lambda}_{k}=\frac{1}{2}
e^{2\phi}p_{k}
\nonumber\\
&&\ddot{\varphi}-\sum_{i=1}^{n}\dot{\lambda}_{i}^{2}=\frac{1}{2}
e^{2\phi}\rho
\label{stcoseqn2}
\end{eqnarray}
(Tseytlin and Vafa, 1992).
The first equation may be used to solve for $\dot{\varphi}$,
so that the second equation becomes
\begin{equation}
\ddot{\lambda_{k}} = \frac{1}{2}p_{k} e^{2\phi}\pm\dot{\lambda}_{k}
\sqrt{\rho e^{2\phi}+\sum_{i=1}^{D}\dot{\lambda}_{i}^{2}-c}
\label{branched}
\end{equation}
With $\rho,p=0$ the solution will be a generalization
of the Kasner solution with
\begin{eqnarray}
\varphi(t) &=& \varphi_{0} - \ln (t_{c} - t) \nonumber \\
\lambda_{i}(t) &=& \lambda_{i,0} + q_{i} \ln (t_{c} - t) \nonumber \\
\sum_{i=1}^{D-1} q_{i}^{2} &=& 1
\label{dilkasn}
\end{eqnarray}
(Tseytlin 1992).
Note that this corresponds to the $+$ branch of the square root
in Equation \pref{branched}.
We wish to match this solution to the Schwarzschild black hole
near the singularity, which as we have argued can
be approximated by a Kasner metric.  Thus, we would start with
the solution to Equation \pref{dilkasn} corresponding to
$\dot{\phi}=0$. At this point we assume that particle production
begins; what happens next depends on the details of the
equation of state.  In particular note that in order to
turn around runaway expansion or contraction it is necessary that
the pressure term become sufficiently large and negative for
rapidly expanding spacetimes and sufficiently large and positive for
rapidly contracting spacetimes.  We will argue in the next section
that string theory gives us exactly this sort of
behavior.

It is worth repeating here a point made by Tseytlin and Vafa (1992).
In Einstein gravity we are used to any energy density at all contributing to
the expansion of the universe, at least for a spatially flat
cosmology where $\dot{\lambda}^{2} = (8\pi G/3) \rho$.
We are arguing that a negative pressure
may in fact provide a drag on the expansion.
The difference is the
presence of the dilaton, which is an embarrassment for phenomenology
but may be a blessing when dealing with singularities.  Note also that
runaway expansion or even a sufficient ammount of (positive) energy
density will cause the dilaton to run to infinite coupling in finite
time, making string loops especially important.  On the other hand,
sufficient contraction and insufficient energy density can cause the
dilaton to run to weak coupling.

\section{String matter as a source for the field equations}

Sanchez and Veneziano (1990) demonstrated that in a large class of
inflationary spacetimes (with a constant dilaton),
certain string modes became unstable
and begin growing with the scale factor.  They argued that
this happens when the string size becomes larger than the horizon
size of the expansion, so that different parts of the string cease
to communicate with each other.  They showed that this occured
potentially with power-law expansion and de Sitter expansion,
and inevitably with superinflationary expansion. Gasperini \etal\ (1991a)
studied the approximate
behavior of large strings in the homogeneous, isotropic metric
\begin{equation}
ds^{2} = \frac{1}{\eta^{2\alpha}}\left(-d\eta^{2} + \sum_{i=1}^{D-1}
(dx^{i})^{2}\right)\ ,
\end{equation}
which describes power-law inflation for $\alpha>1$,
de Sitter inflation for $\alpha=1$, and superinflation for
$0<\alpha<1$.
They found that for sufficiently large strings, the conformal time $\eta$
was proportional to the worldsheet time $\tau$ in conformal
gauge, and that one could expand
the solution to the classical equations of motion and Virasoro
constraints in
a power series expansion in the worldsheet time $\tau$
and thus show
that a consistent lowest-order solution to the equation of motion
plus constraints is
\begin{eqnarray}
&&\vec{X} = \vec{X}(\sigma) + \OO(\tau^{2}) + \mbox{...}
\nonumber\\
&&\eta=\tau\sqrt{\vec{X}^{2}(\sigma)} + \OO(\tau^{2})\ ,
\end{eqnarray}
where $0<\sigma<2\pi$ is the worldsheet spatial coordinate.
Note that in this approximation,  $\vec{X}'\gg\dot{\vec{X}}$,
and $\dot{\eta}\gg\eta'$ as $\tau,\eta\longrightarrow 0$
(where the dot and prime refer to worldsheet time and
space derivatives respectively), which is what one would expect from
a large, slow string.  In a follow-up paper (Gasperini \etal\ 1991b),
they showed that in rapidly contracting spacetimes one finds
unstable behavior for which the worldsheet time is proportional
to the {\em comoving}\ time $t$; the unstable behavior
is characterized by
$\vec{X}'\ll\dot{\vec{X}}$, and $\dot{t}\gg t'$.  (the
former condition came from a term of the form $\tau^{1-\alpha}\vec{X}_{2}$,
where $\alpha<1$, which has a large time derivative
as $\tau\rightarrow 0$.)

Instead of repeating the calculations described above,
we shall instead perform a similar calculation to
find the approximate behavior of very large strings
in a temporarily expanding universe by using the
string size, rather than the temporal distance to an infinite
scale factor, as the expansion parameter.  This will
give us some intuition for the calculations we will perform
in the following two sections.
Consider
an isotropically expanding or contracting universe with metric
\begin{equation}
ds^{2} = C(\eta)\left(-d\eta^{2} + d\vec{X}^{2}\right).
\end{equation}
The classical equations of motion and the classical constraints
in conformal gauge for a string in this spacetime are:
\begin{eqnarray}
&&-\ddot{\eta} + \eta'' -\frac{C_{\eta}}{2C}
\left[\dot{\eta}^{2}-\eta^{'2}+\dot{\vec{X}}^{2}-\vec{X}^{'2}\right]
=0 \nonumber \\
&&-\ddot{\vec{X}}+\vec{X}''-\frac{C_{\eta}}{C}\left[
\dot{\eta}\dot{\vec{X}} - \eta'\vec{X}'\right] = 0 \nonumber \\
&& C \left[\dot{\eta}^{2}+\eta^{'2}-\dot{\vec{X}}^{2}-\vec{X}^{'2}\right]
=0\nonumber\\
&&C\left[-\dot{\eta}\eta' + \dot{\vec{X}}\cdot\vec{X}'\right]=0\ .
\label{fulleom}
\end{eqnarray}
The first two equations are the equations of motion and the last two
are the Virasoro constraints.  We wish to expand these equations
in a power series in the size of the string.  Let $L$ be the
dimensionless length, \ie\ the full length of the string divided by $\lstr$.
Following Gasperini \etal\ (1991a), we assume that the time
dependence of the spatial coordinates is of
lower order.  We thus expand $\vec{X}$ in the series:
\begin{equation}
\vec{X}(\tau,\sigma) = L \vec{X}_{0}(\sigma) +
\sum_{n=1}^{\infty} L^{1-n} \vec{X}_{n}(\tau,\sigma)\ ,
\end{equation}
where the $\vec{X}_{n}$ are all of order $\lstr$.
We must be a little careful about the the worldsheet
time interval we are interested in.  As we will see,
the lowest order solution to the constraints is
$\eta_{0}=\tau L\sqrt{\vec{X}_{0}^{'2}}$;  we will be interested below in
universes which expand for a target space time interval of order $\rho^{-1}$,
where $\lstr L/\rho \gg 1$, so that the worldsheet time interval of
interest will be of order $1/L$.  In order to keep track of
these factors explicitly let us define $s = L\tau$, so that
$s$ will be of order 1 in this expansion.  Then we may expand $\eta$
in the series
\begin{equation}
\eta=\sum_{n=0}^{\infty}\eta_{n}(\tau,\sigma) L^{-n}
\end{equation}
In general we should note that we must expand $C(\eta)$ in
a power series in $1/L$ as well, though this will not show up in the
lowest order equations.  These lowest order equations are
\begin{eqnarray}
&&\ddot{\eta}_{0} + \frac{C_{\eta}(\eta_{0})}{2 C(\eta_{0})}
\left( \dot{\eta}_{0}^{2} + \dot{\vec{X}}_{1}^{2} - \vec{X}_{0}^{'2}
\right) = 0 \nonumber \\
&&\ddot{\vec{X}}_{1} + \frac{C_{\eta}(\eta_{0})}{C(\eta_{0})}
\dot{\eta_{0}}\dot{\vec{X}}_{1} = 0 \nonumber\\
&&\dot{\eta}_{0}^{2}-\dot{\vec{X}}_{1}^{2}-\vec{X}_{0}^{'2} = 0\nonumber\\
&& \dot{\vec{X}}_{1}\cdot\vec{X}_{0}' = 0\ ,
\label{ozeroeom}
\end{eqnarray}
where the dot denotes a derivative with respect to $s$.
One may solve these equations for arbitrary $\vec{X}_{0}(\sigma)$
with $\dot{\vec{X}}_{1} = 0$ (so that we may absorb $\vec{X}_{1}$
into $\vec{X}_{0}$ and thus set $\vec{X}_{1}=0$)
and
\begin{eqnarray}
&&\dot{\eta}_{0}=\sqrt{\vec{X}_{0}^{'2}} \mbox{ or } \nonumber \\
&&\eta_{0}=s\sqrt{\vec{X}_{0}^{'2}}=L\tau\sqrt{\vec{X}_{0}^{'2}}\ .
\label{lowsol}
\end{eqnarray}
The interested reader may check these equations to the next order
and verify that solutions exist given these first order solutions.

Note that in this case, just as in the case of the approach to the
$\eta\rightarrow 0^{-}$ in the work of Gasperini \etal\ (1991),
the comoving string coordinates are frozen to lowest order
(the time derivatives of $\vec{X}$ are of order $1/L$),
so that the proper size of string will grow with the scale factor.  We
shall find that the same happens to the quantum wave functions
below.

If a large number of such unstable strings are
produced as we will claim,
we would like to know their equation of state, \ie\ what the
spacetime stress-energy tensor is for
this collection of strings.  For a single macroscopic string source,
the action for the background fields coupled to this source is
given by Equation \pref{dabhact}.  The
stress tensor is then found by taking the derivative
of this action with respect to the spacetime metric in the
usual manner (Dabholkar and Harvey 1989, Dabholkar \etal\ 1990):
\begin{equation}
T_{\mu\nu}(x) = -\frac{\mu}{\sqrt{G}}\int d^{2}\sigma\sqrt{h}h^{\alpha\beta}
\p_{\alpha}X_{\mu}\p_{\beta}X_{\nu}e^{\gamma\phi}
\delta^{D}\left(X(\tau,\sigma) - x\right))
\label{sourcese}
\end{equation}
Using this ans\"{a}tz for the stress tensor, Gasperini \etal\ (1991b)
show that the approximate equation of state for unstable strings
shrinking with the scale factor in a contracting universe is
\begin{equation}
\rho = \sum_{i=1}^{D-1} p_{i}\ ,
\end{equation}
which is the equation of state for photon-like particles;
for unstable strings growing with the scale factor in an
expanding universe (1991a) they find that
\begin{equation}
\rho = - \sum_{i=1}^{D-1} p_{i}
\end{equation}
This is the sort of behavior we desired in the previous section,
in order to remove the singularity, although for general anisotropic
spacetimes we expect the equation of state to be more complicated.

The ans\"{a}tz shown in Equation \pref{dabhact} for the coupling of
a fundamental string source to the spacetime action
still has the status of a conjecture, although
there is evidence which supports it.
Callan and Khuri (1991), Khuri (1993) and Gauntlett \etal\ (1994)
have found that using this expression and calculating the
low-energy scattering of two such strings using
collective coordinate methods, one can reproduce the scattering
amplitudes found from the low-energy limit of the Virasoro-Shapiro
amplitudes.  Tseytlin (1990) has attempted to derive
Equation \pref{dabhact} by summing up long thin handles on the string
worldsheet (these are essentially wormholes on the worldsheet).
This is a very attractive and probably correct answer but it
requires that one make a connection between the spatial
size of the macroscopic string and the dominance of
``thin handles'' in the integral over the moduli space
of high-genus Riemann surfaces.  We do not understand
how to connect spacetime scales to scales in these moduli spaces,
and believe this to be an interesting problem.  Still,
the intuition that large strings
should be treated as sources of spacetime energy-momentum,
and that this effect should arise from the summation of string
loops describing the interaction of the macroscopic string
with its surroundings, is physically fairly sound; thus,
if the effects of these strings are significant, we know
that the dynamics of string theory causes
certain classes of loop diagrams to become important,
so that we must at least modify the $\beta$-function equations
appropriately. (Note that this is independent of another
reason that loops might become important, namely that the
dilaton may diverge near the singularity.)
We shall comment on this point further
in the conclusion.

\section{String field theory in curved spacetime}

\subsection{How do we evolve the background forward?}

Naturally, we would prefer to understand the
full quantum mechanical behavior of string field theory near
the singularity, where field strengths and quantum fluctuations
are large;
we could then evolve the string field theory wave functional corresponding
to the black hole background from a point where we trust our
classical understanding of the geometry to the dangerous region
near the classical singularity.  Unfortunately this
computation would require a manifestly
background independent quantum formulation
of string field theory,
not to mention a powerful computational control of that theory,
both of which we lack.
Even in the case of ordinary field theory,
if we had a quantum theory of gravity, we would still have to contend
with questions of interpreting the strongly fluctuating spacetime
near the singularity.  We will thus
adopt the standard approach of computing the quantum stress
tensor of matter in a fixed background, in order to
use it as a source term for
the $\beta$-function equations of string theory.

As we stated in the Introduction, this may be a reasonable approach for us.
Since we are given as an adjustable
parameter $\lpl/\lstr$, we will perform
our calculation in a regime where this ratio is small and we can trust
the results of perturbative string theory; then, string physics becomes
important well before quantum fluctuations of spacetime become important.
This is not necessarily the end of the story;
if the background is composed of strings, then at the string scale
there is no clean separation between the background and the string fields we
are quantizing on top of it.

In models with rapidly expanding
directions there is another problem with keeping the background separate;
if gravitons become unstable and grow with the scale factor, then
the background gravitons become macroscopic strings and we must
discuss the background as a soup of macroscopic strings.  Perhaps there
is some way of including this effect self-consistently to get some
sort of modified $\beta$-function equations which are valid in these regimes.
Based on our discussion of macroscopic strings in the previous section, this
would amount to the inclusion of a certain class of loop corrections
to the $\beta$-function equations.  For the present, however,
we will not attack this problem; rather, we will work in a background with a
Hubble volume much larger than one in string units, and
extrapolate the result to $r_{{\rm Hubble}}=\ell_{{\rm string}}$.
If we can do nothing else, we can at least argue that the above issues
must eventually be addressed.

What we would like
to calculate is the renormalized stress-energy tensor for
string fields in a singular background.  This is difficult
in particle field theory, and should be
at least as difficult in string field theory.
As we have mentioned previously,
in string theory we have at our disposal
a few exact conformal field theories
with spacelike singularities (Witten 1991, Giddings \etal\ 1993,1994,
Johnson 1994, Lowe and Strominger 1994, Nappi and Witten 1992).
Since these examples correspond to gauged WZW models we might have some
hope of calculating $\langle T^{\mu\nu}\rangle$
exactly.  However, several problems arise.
The most natural calculation to perform is
that of the one-loop expectation
value
\begin{equation}
\langle T_{{\rm spacetime}}^{\mu\nu}(x) \rangle = \frac{1}{\pi\apr\sqrt{G(x)}}
\langle \int d^{2}\sigma \eta^{\alpha\beta}\p_{\alpha}X^{\mu}\p_{\beta}X^{\nu}
\delta\left(X(\tau, \sigma) - x\right)\rangle
\label{seexp}
\end{equation}
in the first-quantized picture (this expression is derived by taking
the derivative of the first-quantized one-loop partition function -- which
should correspond to the spacetime one-loop effective action for the
background fields (Fradkin and Tseytlin 1985)). For this
calculation we must know
the expectation value for the general off-shell graviton vertex
operator, and so we require an off-shell understanding of the theory
that we do not now possess.  In addition, we must ensure that we are computing
this expectation value \pref{seexp} so that it corresponds to the appropriate
second-quantized vacuum, which means we must find some sort of
prescription for dealing with the poles in the string propagator
(recall that in particle field theory the vacuum is defined
by the Feynman propagator -- see Fulling's book (Fulling 1989)
for a discussion).  Finally, the quantization of strings
propagating on non-compact WZW and gauged WZW models is not well understood.
Given these problems, we will instead make a series of
approximations in order to make the calculation tractable.
First, we will ignore
concerns about the conformal invariance of the background,
which will allow us to work with simplified models of expanding
spacetimes.
Second, we will
work only with the $L_{0}$ constraint on the string wave functions,
rather than the infinite number of equations arising from the
full Virasoro constraints.
Third, we will set the spacetime conformal time $\eta$ to
be a function solely of the worldsheet time $\tau$ in conformal
gauge (otherwise, solving for the string configuration at
a given $\eta$ becomes extremely difficult).
Finally, since
we do not know how to compute the string energy-momentum tensor directly,
we will compute the number of strings produced and from that
calculation determine
the stress-energy tensor.

These approximations should not invalidate our results.
First, we will be looking for semiclassical solutions
to the string wave equation.
As in the computation of the $\beta$-functions using the
background field method, when expanding the first-quantized path
integral of the string around a classical solution,
violations of conformal invariance appear as logarithmic
divergences
in the quadratic fluctuation determinant.
The effects we shall be interested in,
namely the exponential dependence of the production amplitude
on the energy, come from the classical term
$\exp\{(i/\apr)S_{{\rm cl}}\}$ of the wave function.
Second, although we use only the $L_{0}$ constraints rather the full set
of positive-mode Virasoro constraints, imposing
the latter should simply have the effect
of reducing the degrees of freedom of the strings that we produce.
A useful analogy is the computation of Hawking radiation in
QED; if we work in the Lorentz gauge and forget about the
Gupta-Bleuler quantization procedure, we get the same basic physics
out of the calculation, but we find that the calculated Hawking
flux is too large by a factor of $D/(D-2)$, due to the effect of
the unphysical timelike and longitudinal modes of the photon.
A similar story should describe our calculation as well.
Third, although setting $\eta=\eta(\tau)$ is not consistent with
the full Virasoro conditions, as we have seen,
an examination of Equation \pref{fulleom}
shows that this ans\"{a}tz is consistent with the $L_{0}$
constraints.
Finally, although generally the concept of single particles
or single strings is ambiguous in curved spacetime backgrounds,
we will work with a spacetime in which string number does make sense,
namely a temporarily expanding model which becomes
flat Minkowski space in the far past and in the far future.  Here we
may sensibly calculate the number of strings using standard
techniques and then simply multiply the number of strings in a
given quantum mechanical state by the spacetime
energy-momentum tensor of a string in
that state, using the results
described in the previous section, or some modification of them.
Our methods will not be directly extendible
to a calculation of the quantum stress tensor in a spacetime in which the
future is a singularity rather than Minkowski space, but
it will allow us to make some sensible statements about
quantum string field theory in curved spacetime which should apply
to the region we are interested in.

\subsection{Imposing the $L_{0}$ constraints}

We wish to calculate the number of different strings produced just
as we would calculate the number of particles produced: by finding
the solutions to the first-quantized relativistic wave equation
which
are purely positive frequency in the far past and looking for
the coefficients of negative-frequency components in the far future.
For string
theory, the (linearized) wave equation is just the Virasoro condition:
\begin{eqnarray}
        T_{++} + T_{--} &=& G_{\mu\nu}\left(X(\sigma)\right)
                \left(\dot{X^{\mu}}\dot{X^{\nu}} +
        X^{'\mu} X^{'\nu}\right) = 0 \nonumber \\
        T_{++} - T_{--} &=& G_{\mu\nu} X^{'\mu} \dot{X^{\nu}} = 0\ .
\label{virasoro}
\end{eqnarray}
We wish to turn the
Virasoro operators into functional differential operators acting
on wave functionals $\Psi(X(\sigma))$.  In curved space, there
are subtleties even in the case of particle wave functions
due to the necessity of defining the Hilbert space norms
and the path integral measure in a covariant fashion
(DeWitt 1957); we will assume the same story applies when
we derive the string wave operator from the first-quantized
path integral.   The functional differential operator form of \pref{virasoro}
should look like
\begin{eqnarray}
\lefteqn{T_{++} + T_{--} =}   \nonumber \\
&& - \frac{1}{\sqrt{G(X(\sigma))}}
        \frac{\delta}{\delta X^{\mu}(\sigma)}\sqrt{G(X(\sigma))}
        G^{\mu\nu}(X(\sigma))
        \frac{\delta}{\delta X^{\nu}(\sigma)} +   \nonumber \\
&& G_{\mu\nu}(X(\sigma)) X^{'\mu} X^{'\nu} + \xi \RR \nonumber \\
\lefteqn{ T_{++} - T_{--} = X^{'\mu} \frac{\delta}{\delta X^{\mu}}\ ,}
\label{difvir}
\end{eqnarray}
where $\RR$ is the spacetime Ricci scalar and $\xi$ depends on whether
the string is conformally, or otherwise, coupled (in other words, with
what coefficient we multiply a term like $\RR(X(\sigma))$ in the
first-quantized string Lagrangian.)
We prefer to remain agnostic as to how the string
is coupled; at any rate,
in the approximations we shall look at the curvature
term will be of lower order and so we will ignore it.

There are several issues we must deal with before using
Equation \pref{difvir} as a wave operator.  First,
it must be normal ordered.  Secondly, it
represents an infinite number of differential equations, one
for each Fourier mode of each operator.  In general, these Fourier
modes are quite complicated; even if $G$ depends only on the
conformal time $\eta$,
$\eta$ is expandable in Fourier modes and the
metric is generally a nontrivial function.  Finally, it is difficult
to understand and interpret the wave functional in
spacetime since the spacetime time coordinate of the string depends
on the worldsheet spatial coordinate, as it does in Equation
\pref{lowsol}.  With this in mind we will
use only the $\sigma$-independent modes of Equation \pref{difvir}
(in other words, the $L_{0}$ and $\bar{L}_{0}$ constraints),
and we will set $\eta = \eta(\tau)$.

If we write the Fourier expansion of our string modes as
\begin{equation}
X^{\mu}(\tau, \sigma) = \sum_{n=-\infty}^{\infty} X_{n}(\tau) e^{in\sigma}
\end{equation}
and use the reality constraints, we find that $X_{n} = X_{-n}^{*}$.
Let
\begin{equation}
X_{n}^{\mu} = X_{n}^{1\mu} + i X_{n}^{2\mu}\ ,
\end{equation}
so that $X_{-n} = X^{1}_{n} - i X^{2}_{n}$.  We then find that the $L_{0}$
part of \pref{difvir} is, for the homogeneous spacetimes we are interested in,
\begin{eqnarray}
\lefteqn{\left\{ \frac{1}{\sqrt{G(\eta)}} \frac{\p}{\p\eta}
\sqrt{G(\eta)} G^{\eta\eta}
\frac{\p}{\p\eta} - \sum_{k=1}^{d-1} \sum_{n=1}^{\infty}
G^{kk}(\eta) \left((\nabla_{k,n}^{(1)})^{2} +
(\nabla_{k,n}^{(2)})^{2}\right)\right. - }   \nonumber \\
&&\left.\phantom{\left\{\right.}\sum_{k=1}^{d-1} G^{kk}(\eta)
\left(\frac{\p}{\p X^{k}_{0}}\right)^{2} +
\right.\nonumber\\
&&\left.\sum_{k=1}^{d-1}\sum_{n=1}^{\infty} n^{2} \left((X_{1,n}^{k})^{2} +
(X_{1,n}^{k})^{2}\right)^{2} - \xi \RR \right\} \Psi\left[\eta,
\{ X_{n} \}\right] = 0  \nonumber \\
\lefteqn{\left\{ \sum_{k=1}^{d-1} \sum_{n=1}^{\infty} n \left( X_{1,n}^{k}
\frac{\p}{\p X_{2,n}^{k}} - X_{2,n}^{k} \frac{\p}{\p X_{1,n}^{k}} \right)
\right\} \Psi\left[ \eta, \{ X_{n} \}\right] = 0\ ,}
\label{lzero1}
\end{eqnarray}
(where we take the metric to be diagonal).
Here $\nabla^{(i)}_{\mu,n} = \p/\p X^{(i)\mu}_{n}$.
Imposing the final constraint in Equation \pref{lzero1}
becomes much simpler if we set
\begin{equation}
X_{n} = R_{n} e^{i \phi_{n}}\ ,
\end{equation}
as Equation \pref{lzero1} becomes:
\begin{eqnarray}
\lefteqn{\left\{ \frac{1}{\sqrt{G}} \frac{\p}{\p\eta} \sqrt{G} G^{\eta\eta}
\frac{\p}{\p\eta} - \right.}   \nonumber \\
&& \left.\phantom{\left\{\right.}\sum_{k=1}^{d-1} \sum_{n=1}^{\infty}
G^{kk} \left( \frac{1}{R_{n}^{k}} \frac{\p}{\p R_{n}^{k}}
R_{n}^{k} \frac{\p}{\p R_{n}^{k}} + \frac{1}{\left(R_{n}^{k}\right)^{2}}
\left(\frac{\p}{\p \phi_{n}^{k}}\right)^{2}\right)\ - \right. \nonumber \\
&&\left.\phantom{\left\{\right.}
\sum_{k=1}^{d-1} \sum_{n=1}^{\infty} n^{2} \left(R_{n}^{k}\right)^{2}
- \xi \RR \right\} \Psi\left[ \eta, \{ X_{n} \} \right] = 0 \nonumber \\
\lefteqn{\sum_{k=1}^{d-1} \sum_{n=1}^{\infty} n \frac{\p}{\p \phi_{n}^{k}}
\Psi\left[ \eta, \{ R_{n}, \phi_{n} \}\right] = 0\ .}
\label{lzero2}
\end{eqnarray}
Note in particular that these constraints commute with $\p/\p\phi_{n}^{k}$
for all $n$,$k$, so that we may let
\begin{equation}
\Psi\left[\{R,\phi\}\right] = \exp\{i \sum_{k,n}
l_{k,n} \phi_{n}^{k}\}
\Psi\left[\{R\}\right]\ ,
\end{equation}
and thus make the substitution
\begin{equation}
\frac{\p}{\p \phi_{n}^{k}} \longrightarrow i l_{k,n}\ .
\end{equation}

It will be useful at this point to write down the stationary states
of \pref{lzero2} in flat space.  We will add a term
${\cal E}_{0}^{2}\Psi$ to this equation in order to explicitly subtract off
the zero-point energy: this will include the usual tachyon shift
($-m_{0}^{2}$)
if we are working with the bosonic string.
In flat space the equation clearly separates
into an infinite number of 2-dimensional harmonic oscillators,
one for each $(k,n)$, with the single constraint that
\begin{equation}
\sum_{k=1}^{d-1}\sum_{n=1}^{\infty} n l_{k,n} = 0\ .
\label{spinconstr}
\end{equation}
We also get a Klein-Gordon equation for the $X_{0}$ wave functions.
The final wave function may then be written in terms of associated
Laguerre polynomials.  If we leave the scale factor $\Omega$ of the
metric in explicitly, we find that
\begin{eqnarray}
\Psi_{\{m_{k,n},l_{k,n}\}} &=& e^{-i E \eta + i \vec{q}\cdot\vec{X}_{0}}
\times \nonumber \\
&&\prod_{k,n_{k}} e^{-n_{k}\Omega(R_{n}^{k})^{2}/2}
\left(n\Omega(R_{n}^{k})^{2}\right)^{l_{k,n}/2}
L_{m_{k,n}}^{l_{k,n}}\left(n\Omega(R_{n}^{k})^{2}\right)e^{i l_{n,k}
\phi_{n}^{k}}\ .
\nonumber \\
\label{flatfunc}
\end{eqnarray}
Here
\begin{eqnarray}
E^{2} &=& \vec{q}^{\ 2} + \sum_{n,k}4n\Omega\left(m_{k,n}
+\frac{l_{k,n} + 1}{2}\right) - {\cal E}_{0}^{2} \nonumber \\
&=& \sum_{n,k}4n\Omega\left(m_{k,n}+\frac{1}{2}\right) - {\cal E}_{0}^{2}\ ,
\label{flatenergy}
\end{eqnarray}
and the final line in \pref{flatenergy} comes from \pref{spinconstr}.
Note that $\Omega$ acts as the frequency of the harmonic oscillator,
and the only difference between this system and a product of nonrelativistic
harmonic oscillators is that $E$ gets replaced by $E^{2}$.
As usual, $\EE_{0}$ is defined such that
\begin{equation}
E^{2} = \vec{q}^{\ 2} + \sum_{n,k}4nm_{k,n} - m_{0}^{2}\ .
\end{equation}

\subsection{Approximations for cosmological models}

Let
\begin{equation}
ds^{2} = - dt^{2} + \sum_{k=1}^{d-1} a_{k}(t) dx^{2}_{k}\ .
\end{equation}
Putting this metric into Equation \pref{lzero2}, we find that
the scale factors $a(t)$ act as time-dependent masses for the
different 2D harmonic oscillators.
In this case, Equation \pref{lzero2} is not separable, and no
interesting scale factors yielding exact solutions are known to
us. (This is in contrast to the non-relativistic harmonic oscillator
with a time-dependent mass and/or frequency, where one can
reduce the quantum problem to the classical problem by
introducing a Gaussian wave packet -- essentially a coherent-state
wavepacket -- with time-dependent coefficients (Peremolov and Popov 1969).
There is no solution in the relativistic case as there
are no analogs of coherent states: even with a time-independent
mass and frequency, the relativistic dispersion relation causes the
wave packet to spread.) We will instead examine Equation \pref{lzero2}
in certain limiting cases.

\subsubsection{The adiabatic limit}

The simplest limit to study is the limit of an adiabatically
changing universe,
where $a_{k} = a_{k}(\rho t)$,
$1/\rho \ll \lstr$ (note we have so far set $\lstr = 1$).  Let
us work to lowest adiabatic order; since the curvature scalar is
second order in derivatives of the metric, and thus $\OO(\rho^{2})$,
we will ignore this term for the present.  To this order, the
oscillator number should be an adiabatic invariant, so we
use the following variant of the flat space energy eigenfunctions:
\begin{eqnarray}
\lefteqn{\Psi^{{\rm ad.}}_{m_{k,n}, l_{k,n}}\left[\{R, \phi\}\right] = }
\nonumber   \\
&& \prod_{k,n_{k}} e^{-n_{k}a_{k}(\eta)(R_{n}^{k})^{2}/2}
\left(n a_{k}(\eta)(R_{n}^{k})^{2}\right)^{l_{k,n}/2}
L_{m_{k,n}}^{l_{k,n}}\left(n a_{k}(\eta)(R_{n}^{k})^{2}\right)
e^{i l_{k,n} \phi_{n}^{k}}\ . \nonumber \\
\end{eqnarray}
If we now expand the total wave function in this basis using time
dependent coefficients,
\begin{equation}
\Psi = \sum_{\{m_{k,n},l_{k,n}\}} c_{\{m_{k,n},l_{k,n}\}}(t)
\Psi^{{\rm ad.}}_{\{m_{k,n},l_{k,n}\}}\ ,
\label{adexpand}
\end{equation}
then applying Equation \pref{lzero2} to this wave function will
give us a coupled set of differential equations for the coefficients
$c_{\{m_{k,n},l_{k,n}\}}(t)$, subject to the initial
conditions appropriate to the question one is asking.

We would like to compute the probability of particle production.
As usual, we first need to define the positive frequency modes
of the string field in the far past and future.  The standard
definition is the
definition of adiabatic positive frequency modes (see Birrell and
Davies (1982) for an explanation and references): we
ask that the modes we define as ``positive frequency''
be those that match exactly onto positive frequency
WKB solutions in the infinite past,
\begin{equation}
\Psi^{{\rm ad.}}_{\{m_{k,n},l_{k,n}\}} \frac{1}{\sqrt{E(t)}}
\exp\left\{ -i \int_{-\infty}^{t} dt' E(t')\right\}\ ,
\label{adevol}
\end{equation}
where for our spacetime
\begin{equation}
E(\eta) = \sqrt{ \sum_{k} \frac{(q^{k})^{2}}{a_{k}(t)}
+ \sum_{n} 4 n m_{k,n} + \xi \RR}\ .
\label{enevol}
\end{equation}
Ideally, we would find the appropriate exact wave function that matches onto
this WKB solution and look for the negative frequency components in
the usual manner: in lieu of this, we would like a WKB approximation
for the wave function at all times.
Two problems arise, however.  The first problem is
that the negative frequency
components will not be present at any order in the perturbation
series one would get from plugging \pref{adexpand} into \pref{lzero2}.
The reason is identical to the reason one does not find above-barrier
reflection in semiclassical non-relativistic quantum mechanics to
any perturbative order in $\hbar$.  The second problem is that the terms
in the equations for the coefficients $c_{m,l}(t)$ in
Equation \pref{adexpand}
which are higher order in $\rho$ come from applying the
time derivatives to $\Psi^{{\rm ad.}}$, which depend implicitly
on time through $a_{k}$.  The reader may easily verify that
these terms are also multiplied by factors of $m_{k,n}$ and
higher powers of $m$, because of the recursion relations
for Laguerre polynomials.  Thus if $m_{k,l}\rho \gg 1$, the
adiabatic approximation breaks down.  We will treat this case below.

In the case that the adiabatic approximation holds, there is still
a trick for extracting the negative frequency part of the wave function
in the far future.  Audretsch (1979) has pointed out, for scalar particles
in curved spacetime, that
the particle production calculation is almost identical
to the calculation of
above-barrier reflection in nonrelativistic quantum mechanics, since
one can generally (for homogeneous spacetimes) separate the
time and space parts of the wave equation and write the former as
\begin{equation}
\frac{\p^{2}}{\p t^{2}} f + \omega(t)^{2} f = 0\ .
\end{equation}
The difference is that in above-barrier reflection, we want the
incident component of the wave
function to be normalized to unity and the
transmitted wave function to be purely outgoing.  Then,
if the amplitude of the transmitted wave is $T$ and the amplitude of the
reflected wave is $R$ (assuming that only one value of $|\omega_{t=\infty}|$
contributes
to the final wave function, though the discussion may
be easily generalized)
probability conservation requires that
\begin{equation}
|T|^{2} + |R|^{2} = 1\ .
\end{equation}
On the other hand, in the case of non-static spacetimes,
we wish the wave function in the far past
to be purely positive frequency.  If this incoming wave function is
normalized to unity, and the coefficients of the positive
and negative frequency components of the outgoing
wave function are denoted $\alpha$ and $\beta$ (again we assume that only one
magnitude of the frequency contributes to the final wave function),
probability conservation
requires that
\begin{equation}
1 + |\beta|^{2}= |\alpha|^{2}\ .
\end{equation}
If we write  $\omega = \sqrt{2(E-V(t))}$ the analogy is obvious.
In general, $\omega$ will vanish in the upper half complex $t$-plane
at some point (or set of points) $t_{0}$, and also at $t_{0}^{*}$.  The
square of the reflection coefficient in the WKB approximation
has been derived by Pokrovskii and Khalatnikov (1961):
\begin{equation}
|R|^{2} = \exp\left\{ - 2\ {\rm Im} \int_{t_{0}^{*}}^{t_{0}}
dt\ \omega(t)\right\}\ .
\label{wkbrefl}
\end{equation}
In our case, if we substitute $E$ for $\omega$,
$|R|^{2}$ becomes the square of the Bogolubov coefficient multiplying the
negative frequency WKB solution in the far future (Audretsch 1979).

\figloc1{Complex time paths.  $t_{0}$ and $t_{0}^{*}$ are the classical turning
points of the system.  Curve C denotes a path which moves around the
square root branch point, while staying far enough away for the
adiabatic or semiclassical approximation to hold.  For the
purposes of computing the production amplitude the integral
$\int dt E$ may be performed along curve C'.}

Before we continue, let us describe a more heuristic
derivation of Equation \pref{wkbrefl} (see also
Landau and Lifshitz (1977) and Migdal (1977), from which this
discussion was taken).  As we have
seen, the time dependence of our system has the form of Equation
\pref{adevol}.  The expression for $E$ in equation \pref{enevol}
contains square root branch cuts in the complex $t$ plane (see Figure 1).
In order to detect the part of the wave function with negative energy,
we should go around the end of the branch cut in $E$.  If
we stay far enough away from this endpoint, then the
adiabatic approximation will remain valid.
(The approximation will
break down near the turning point, where $E=0$, since one
requires for the approximation that the time variation of
the wave function be more rapid than the time variation of
the background.)  Along the contour,
labeled $C$ in Figure 1, the integral $\int dt E$ will pick up an
imaginary component, which will give us the real part
of the Bogolubov coefficient $\beta$.  This real part will not change if we
move the points $a1$,$a2$, where the contour $C$ leaves
the real axis, adjacent each side of the branch cut;
so long as $C$ does not enclose any singularities other
than the branch cut, we may then shrink the contour down to the
curve $C'$ which
surrounds the branch cut as shown.  A little manipulation of
the integral should convince the reader that
the imaginary part of the integral in Equation
\pref{wkbrefl} is then the correct result for the
real part of the phase picked up when
traveling on contour $C$.  Since $\exp\{-E/\rho\}$ will be
small in the adiabatic limit, the correction for
the correct normalization of the wave function will be
even smaller, and we may identify the expression shown in
equation \pref{wkbrefl} with $|\beta|^{2}$.  This
discussion can be easily generalized to systems with several turning
points in the complex plane (Pokrovskii and Khalatnikov 1961,
Landau and Lifshitz 1977, Migdal 1977, Audretsch 1979).

\figloc2{Pair production of strings.  Here we depict the configuration
of the classical string as it passes around the branch point in
the complex $t$ plane.  One may interpret this as a pair
production process beginning at $t_{0}$; the string which reaches
the real axis on the left of the branch cut is positive frequency and the
string which reaches the right side of the branch cut is negative frequency.}

One can envision the pair production process via the classical
trajectory of the string on $C'$.
Since the energy is small near the turning point,
the size of the string will be small as well.  The
classical trajectory of the string describes two
zero-size, zero-energy strings being produced at $t_{0}$,
and growing as they travel in imaginary time along opposite sides
of the branch cut (see Figure 2).

\subsubsection{The semiclassical limit}

In the limit that the occupation numbers $m_{k,n}$
are large, the adiabatic approximation breaks down and
the semiclassical approximation is appropriate.  Let us write
the wave function as
\begin{equation}
\Psi = \sum_{j}D_{j}(t, \{X_{n}\}) e^{\frac{i}{\alpha'} S_{j}(t, \{x_{n}\})}
\ .
\label{semicwf}
\end{equation}
In the semiclassical limit, where the frequencies (excitation numbers
and center of mass momentum) are large, $S$ will solve the
Hamilton-Jacobi equation
\begin{eqnarray}
&&-\left(\frac{\p S}{\p t}\right)^{2}
        + \sum_{k}\frac{(q^{k})^{2}}{a_{k}} +
\sum_{k,n}\frac{1}{a_{k}(t)}\left[
\left(\frac{\p S}{\p R_{n}^{k}}\right)^{2} +
\left(\frac{1}{R_{k}^{n}}\frac{\p S}{\p \phi_{n}^{k}}\right)^{2} \right]
+ a_{k}(t) \left(R_{n}^{k}\right)^{2} \nonumber \\
&&\phantom{-\left(\frac{\p S}{\p t}\right)^{2}} + \xi \RR - m_{0}^{2} = 0\ ,
\label{lzerohj}
\end{eqnarray}
where $S$ defines the momentum as a function of position via
$p_{n} = \p S/\p q_{n}$,
$p$ is a multivalued function defined by a curve $\CC$ in
phase space (see Figure 3) and the sum in \pref{semicwf} is the sum over
the branches of the function $p(q)$ defined by $\CC$.
If we foliate phase space with curves $\CC = \CC(\PP)$, then we may write
$D_{j}$ as the square root
of the Van Vleck determinant
\begin{equation}
D_{j} = \mbox{constant}\ \times\  \sqrt{
\left| \frac{\p^{2} S_{j}}{\p q_{l}\p\PP_{k}} \right|} \ .
\end{equation}
(This geometric picture is taken from Berry and Balasz (1979) and is
due to Maslov (1965).  See Berry (1981) for further review and references.)
In general the condition for the semiclassical approximation to
hold is that:
\begin{eqnarray}
\frac{\p^{2}S}{\p q_{n}^{2}}&\ll&\left(\frac{\p S}{\p q_{n}}\right)^{2}
\label{wkbcon1} \\
\frac{\p^{2}S}{\p t^{2}}&\ll&\left(\frac{\p S}{\p t}\right)^{2}\ .
\label{wkbcon2}
\end{eqnarray}
This is the usual criteria for the validity of the WKB approximation.

\figloc3{Curves $p(q)$ in phase space.  The n-dimensional
surface $\CC_{\eta}(\PP)$ defines
a multivalued function $p_{k}(\{q_{l}\})$ at time $\eta$ in a 2n-dimensional
phase space.  $\PP$ labels a
family of n-dimensional curves which foliate phase space.
$\CC_{\eta 1}(\PP_{1})$ is found by evolving each point on
$\CC_{\eta}(\PP_{1})$ from time $\eta$ to time $\eta'$ via
Hamilton's equations.}

Note that in general the Hamilton-Jacobi equation for a relativistic
system should look like
\begin{equation}
\frac{\p S}{\p\tau} +
\HH\left(t,x,E=\frac{\p S}{\p t}, p=\frac{\p S}{\p x}, \tau\right) = 0\ ,
\end{equation}
if we are to treat space and time on an equal footing.
$\tau$ should be an arbitrary parameter describing the
world line of this system, \ie\ the $\tau$ dependence of the system
should be trivial.  Thus,
we arrive at Equation \pref{lzerohj}, where
\begin{eqnarray}
\lefteqn{\HH \left(t, E, R_{n}^{k}, p_{R,n}^{k}, \phi_{n}^{k},
p_{\phi,n}^{k}\right) = }    \nonumber \\
&& - E^{2} + \sum_{k} \frac{(q^{k})^{2}}{a_{k}} + \sum_{k,n} \left\{
\frac{(p_{R,n}^{k})^{2}}{a_{k}(t)} +
\frac{(p_{\phi,n}^{k})^{2}}{(R_{n}^{k})^{2}a_{k}(t)} +
a_{k}(t) (R_{n}^{k})^{2}\right\}\nonumber\\
&& + \xi\RR - m_{0}^{2}= 0\ .
\label{stringhj}
\end{eqnarray}
The simplest way forward is to simply
solve for the constraints, by letting $E$, the generator of $t$
translations, become the Hamiltonian and using
Equation \pref{stringhj} to write
\begin{eqnarray}
\lefteqn{E\left(R,p_{R},\phi,p_{\phi},t\right)}     \nonumber\\
&&= \sqrt{\sum_{k} \frac{(q^{k})^{2}}{a_{k}} + \sum_{k,n}
\left\{ \frac{(p_{R,n}^{k})^{2}}{a_{k}(t)} +
\frac{(p_{\phi,n}^{k})^{2}}{(R_{n}^{k})^{2}a_{k}(t)} +
a_{k}(t) (R_{n}^{k})^{2}\right\} +(\xi\RR - m_{0}^{2}} = 0\ . \nonumber \\
\label{gfham}
\end{eqnarray}
This also comes from putting the term
$(\p S/\p t)^{2}$ on the right hand side of Equation
\pref{lzerohj}
and taking the square root of the equation.

Now we must solve the Hamilton-Jacobi equation
\begin{equation}
\frac{\p S}{\p t} + E = 0\ .
\end{equation}
We describe the solution as follows (this
discussion is patterned after that in Berry and Balazs (1979),
though our representation for $S$ is different).
At some initial time, say $t=0$, we start with
a curve $\CC_{0}$ in phase space which defines the desired initial
wave function via \pref{semicwf}.  This curve will evolve forward in time,
as one can see by taking each point on the curve as an initial condition
for Hamilton's equations, to a curve $\CC_{t}$ at time $t$.  This
will define a family of functions $p(q,t)$.  Assuming
no new caustics in $p(q,t)$ develop, the solution to the Hamilton-Jacobi
equation can be written, to within a constant, as:
\begin{equation}
S = \int_{q_{n,0}}^{q_{n}} p_{n}\left(q,t_{0}\right)dq_{n} -
\int_{t_{0}}^{t} E\left(q_{n}, p_{n}(q_{n}, t'), t'\right) dt'\ .
\label{genhjsoln}
\end{equation}
Here the $q_{(n,0)}$ are arbitrary constants.  A straightforward
application of Hamilton's equations confirms that
\begin{equation}
\frac{\p S}{\p q} = p(q),
\end{equation}
and that $S$ as defined in Equation \pref{genhjsoln} satisfies the
Hamilton-Jacobi equations.

One may compute string production in the same fashion as one
would in the adiabatic approximation: here one would search for turning
points in the complex time plane for $p_{n}$ and $E$.  To see this
procedure in
an explicit model we turn to the next section.

\section{An isotropically expanding model}

We would now like to examine a model from which we may extract
a few results.  For the reasons laid out in the beginning of the
previous section, we would like to examine a model in which
the far past and far future correspond to Minkowski space and there is a
finite period of expansion near $\eta=0$.  At present the one case
we have found to be tractable is the case of isotropic expansion.
In this case we will work in conformal time:
\begin{equation}
ds^{2} = C(\eta) \left( -d\eta^{2} + \sum_{k} (dx^{k})^{2}\right)\ .
\end{equation}
A simple scale factor with the desired properties is
$C(\eta) = A + B \tanh \rho \eta$, where $A-B = 1$ and $A + B = \Omega$,
the final scale factor.  The conformally coupled Klein-Gordon equation in
a two-dimensional spacetime with this scale factor has been solved
exactly and examined
by Bernard and Duncan (1977); the time part of the equation is
identical to the Schr\"{o}dinger equation in a special case
of the Eckhardt-Sauter potential (Eckhardt 1930, Sauter 1932).
It will be useful as a warmup exercise
to compute the production probability for particles
in this spacetime
using the WKB approximation, since we
have an exact solution to compare it to.  After some work we will find
the semiclassical result for string production to be nearly identical.

\subsection{Review of the Klein-Gordon equation}

For a particle with mass $m$, we are looking for solutions which look in the
far past and far future like:
\begin{eqnarray}
t\longrightarrow -\infty : && \phi(x,t) \sim e^{-i\omega_{-}t + i q x}
\nonumber \\
t\longrightarrow \infty : && \phi(x,t) \sim \alpha_{q} e^{-i\omega_{+} t
+ iqx} + \beta_{q} e^{i\omega_{+}t + iqx}\ ,
\label{asymp}
\end{eqnarray}
where
\begin{equation}
\omega_{\pm}^{2} = q^{2} + (A\pm B) m^{2}\ .
\end{equation}
Bernard and Duncan (1977) found exact solutions to the Klein-Gordon
equation (in the form of hypergeometric functions) which match to
\pref{asymp}.  The exact expression for the production probability is
\begin{equation}
|\beta_{k}|^{2} = \frac{\sinh^{2}
\left(\pi(\omega_{+}-\omega_{-})/2\rho\right)}
{\sinh\left(\pi\omega_{-}/\rho\right)\sinh\left(\pi\omega_{+}/\rho\right))}\ .
\end{equation}
In the limit of small $\rho$ (or large energies and energy differences),
this expression becomes
\begin{equation}
|\beta_{k}|^{2} = e^{-2\pi\omega_{-}/\rho}\ .
\label{refasymp}
\end{equation}

Using the WKB approximation, we find that Equation \pref{wkbrefl} becomes
\begin{equation}
|\beta_{k}|^{2} = \exp\left\{ - 2 {\rm Im} \int_{\eta_{0}^{*}}^{\eta_{0}}
\sqrt{ k^{2} + (A + B \tanh \rho t) m^{2}}\right\}\ ,
\label{eckwkb}
\end{equation}
where
\begin{equation}
\eta_{0} = \frac{i\pi}{2\rho} + \frac{1}{2\rho} \ln \left(
\frac{\omega_{-}}{\omega_{+}}\right)\ .
\end{equation}
If we make the substitution
\begin{equation}
\eta = \ln \left(\frac{\omega_{-}}{\omega_{+}}\right)+\frac{\ln z}{\rho}\ ,
\end{equation}
then the integral in Equation \pref{eckwkb} becomes
\begin{equation}
\frac{\omega_{-}}{2\rho} \oint_{|z|=1} \frac{dz}{z}
\sqrt{\frac{z + 1}{z + \frac{\omega_{+}}{\omega_{-}}}}
= \frac{i\pi \omega_{-}}{\rho}\ .
\end{equation}
Plugging this into Equation
\pref{eckwkb} we find that this approximation reproduces
Equation \pref{refasymp}.

\subsection{Strings in an expanding universe}

For states for which the adiabatic approximation holds, it is easy to
see that if we ignore $(\xi\RR- m_{0}^{2})$
in the equation, we may use the approximations
described to arrive at Equation
\pref{refasymp}.  However, states for which this
approximation holds are not generic at sufficiently high mass levels.
String theory contains an infinite
number of states with a multiplicity growing as the mass.  For a given
level $M=(\mbox{mass})^{2}$ of string excitation,
states corresponding to high Fourier
components $K$ of the string excited to low oscillator levels
$m_{K}$, such that
$m_{K}\rho \ll 1$, can still be treated with the adiabatic method.
But in general the typical size of the string
will be large, reflecting a large number of Fourier components
excited to a high oscillator level.
We will find that while some distributions of
the energy for a given $M$ are intractable even in the semiclassical
limit, i.e. those with a few Fourier components highly excited,
a configuration with $N \gg 1$ Fourier modes excited
may be treated in a $1/N$ expansion.  The latter type of
configuration should be
fairly generic for sufficiently large mass; we expect that most
configurations at a given total oscillator level $n$ will
have their energy partitioned
among many oscillator modes.

In conformal coordinates, the scale
factor appears as a time-dependent frequency rather than a time-dependent
mass; thus, the energy (the generator of $\eta$ translations) is
\begin{equation}
E=\sqrt{\vec{q}^{\ 2}+C(\eta)(\xi\RR - m_{0}^{2})
+\sum_{k,n}(p_{R,n}^{k})^{2}+(nC(\eta)R_{n}^{k})^{2}+
\left(\frac{p_{\phi,n}^{k}}{R_{n}^{k}}\right)^{2}}\ .
\end{equation}
Hamilton's equations are:
\begin{eqnarray}
&&\dot{R}_{n}^{k} = \frac{p_{R,n}^{k}}{E} \nonumber \\
&&\dot{p}_{R,n}^{k} = \left(\frac{(p_{\phi,n}^{k})^{2}}{(R_{n}^{k})^{2}}
- (nC(\eta))^{2} R_{n}^{k}\right)/E \nonumber \\
&&\dot{\phi}_{n}^{k} = \frac{p_{\phi,n}^{k}}{E (R_{n}^{k})^{2}}
\nonumber \\
&&\dot{p}_{\phi,n}^{k} = 0\ .
\label{hameqn}
\end{eqnarray}
(There is also the remaining constraint $\sum_{k,n} nl_{k,n}=0$.)
These equations may be converted into equations of motion for $R$ and $\phi$:
\begin{eqnarray}
&&\ddot{R_{n}^{k}}+\left(\frac{nC(\eta)}{E}\right)^{2} R_{n}^{k}
- \frac{(p_{\phi,n}^{k})^{2}}{E^{2}(R_{n}^{k})^{3}} +
\frac{\dot{E}}{E} \dot{R_{n}^{k}} = 0 \nonumber \\
&&\ddot{\phi_{n}^{k}} + \frac{2p_{\phi,n}^{k}\dot{R_{n}^{k}}}{(R_{n}^{k})^{3}}
+ \frac{\dot{E}}{E} \dot{\phi_{n}^{k}} = 0\ .
\label{eom}
\end{eqnarray}
Note that without the $\dot{E}/E$ terms, Equation \pref{eom}
looks like the equations
for a harmonic oscillator with frequency
\begin{equation}
\bar{\omega}_{n} = \frac{n C(\eta)}{E}\ .
\label{redfreq}
\end{equation}

If a single oscillator is excited,
Equation \pref{eom} becomes quite difficult
to solve in the limit $E\rho/\omega \gg 1$.  The $\dot{E}/E$ term is of lower
adiabatic order, but the reduced frequency is smaller still.  The
period of oscillation in spacetime time is much longer than the
time scale $\rho^{-1}$.  Whether $\omega R/E$ is of order 1, of order $\rho$,
or of order $1/E$, and thus whether the velocity- or position-dependent force
dominates, will depend on where in the potential the
oscillator term is.  At present, we have no ideas to offer for
a solution in this case.

However, as we have stated above,
most of the states at a given large mass will be
in oscillators with many Fourier modes
excited -- call this number $N$.
We will attempt to solve for the classical equations to lowest order
in $1/N$.  This is similar to the expansion in the
size of the string we presented two sections ago; the
string in such a state will be large and slow.  We will assume
that for a given oscillator mode $(k,n)$ with
energy $E_{k,n}$, $E_{k,n}\rho/\omega \gg 1$ and
$N \gg 1$; thus $E \gg E_{k,n}$ for any $(k,n)$.

Using Hamilton's equations we can see that
\begin{equation}
\frac{\dot{E}}{E} = \frac{\dot{C}}{C}
\frac{\sum_{k,n} n^{2} C^{2} (R_{n}^{k})^{2}}{E^{2}}\ .
\label{edovere}
\end{equation}
The expression multiplying $\dot{C}/C$ is $\OO(1)$ both in
$1/N$ and in $\rho$.  The position-dependent terms in \pref{eom}
are of order $1/N$.  Even though $N\rho \gg 1$, the
oscillator terms will dominate for most of the life of the
universe, where the $\tanh(\rho\eta)$ term in $C$ is exponentially
suppressed and so not changing much.
More specifically, note that
\begin{eqnarray}
\frac{\dot{C}}{C} &=& \frac{B\rho\ \rm{sech}^{2}\rho\eta}{A+B\tanh\rho\eta}
\nonumber \\
&\sim& B\rho e^{-2\rho|\eta|}
\mbox{     for $\eta\longrightarrow\pm\infty$}\ .
\end{eqnarray}
Thus, noting that Hamilton's equations tell us that
$\dot{R}\sim\OO(1/\sqrt{N})$, so long as
\begin{equation}
|\eta| > \frac{1}{2\rho} \ln\left\{|B|\rho\sqrt{N}\right\} = \eta_{f}\ ,
\end{equation}
the position dependent forces dominate and the system behaves
as a collection of simple harmonic oscillators.  Since
\begin{equation}
\dot{E}\sim\frac{\dot{\omega}}{\omega}\OO\left(\sqrt{N}\right)\ ,
\end{equation}
for $|\eta|>\eta_{f}$ the change in $E$ (or in $\omega$) becomes
very small exponentially rapidly, and the
system behaves as set of oscillators with frequency
\begin{equation}
\bar{\omega}_{\pm} = \frac{nC(\pm\infty)}{E(\pm\infty)}\ .
\end{equation}
Note that the factor of $1/E$ means that the potential is
quite shallow and in general each Fourier mode will have
a spatial extent of order $\sqrt{N}$: the string will be
quite large, as promised.
For $|\eta|<\eta_{f}$, the velocity dependent term dominates.
Keeping only these terms and the second derivative in \pref{eom},
we find that
\begin{eqnarray}
\dot{R_{n}^{k}}&=& \frac{\alpha_{R,n}^{k}}{E} \nonumber \\
\dot{\phi_{n}^{k}}&=& \frac{\alpha_{\phi,n}^{k}}{E}\ ,
\end{eqnarray}
where the $\alpha$ are integration constants and will be fixed by
matching this solution to the $|\eta|>\eta_{f}$ region.
Thus the velocities are $\OO(1/\sqrt{N})$: we shall call
this region the ``frozen region'' (thus the suffix in $\eta_{f}$).
The changes in these coordinates are:
\begin{equation}
\delta (R,\phi) \sim 2 \eta_{f} (\dot{R},\dot{\phi}) \sim
\frac{1}{\rho\sqrt{N}}\ln\left\{|B|\rho\sqrt{N}\right\}\ .
\end{equation}
As long as $\rho\sqrt{N}\gg 1$, the change in the coordinate
positions is small; therefore, we will approximate the string in
this region as being fixed in comoving coordinates, so
that its {\it proper} size grows with the scale factor, as
we saw two sections ago.

Physically, the picture that we have is that strings
which are larger than a horizon radius during expansion
become ``frozen'' and expand with the scale factor without
oscillating, as different parts of the string are no longer in
causal contact with each other.
The entire history of the string can be divided into three regions:
\begin{itemize}
\item{$-\infty<\eta<\-\eta_{f}$. Strings oscillate
freely and energy is conserved.  Here we may expand the
string wave function in stationary states of the initial Hamiltonian.}
\item{$-\eta_{f}<\eta<\eta_{f}$ (the
``frozen'' region.)  Classically, strings larger than
the horizon size do not oscillate coherently since different regions
of the string are causally disconnected from each other.
Instead, the string grows with the scale factor.  Equivalently,
in conformal coordinates the string wave function is frozen as
the harmonic oscillator potentials decrease in width.}
\item{$\eta_{f}<\eta<\infty$.  The wave function at
$\eta=\eta_{f}$ is
no longer a stationary state of the final Hamiltonian:
we may, within our approximation, decompose it
into eigenstates of the final Hamiltonian,
just as in the sudden approximation.}
\end{itemize}
In the language of Berry and Balazs (1979),
if we are looking for a wave function corresponding to
an energy eigenstate with energy $E_{-}$ in the far past, we wish to
start with a curve $\CC_{-\infty}$ in phase space
corresponding to the torus defined by the equation $E=E_{-}$
in phase space.  In this region the energy of each oscillator,
$E_{k,n}$, is a constant of motion, so that
\begin{equation}
p_{R,n}^{k} = \pm \sqrt{(E_{n}^{k})^{2} - \bar{\omega}_{-}^{2}
(R_{n}^{k})^{2} - \left(\frac{l_{n}^{k}}{R_{n}^{k}}\right)^{2}}\ .
\label{pofq}
\end{equation}
The branch of the square root changes when we reach a caustic
in $p(q)$ and, as stated, we must sum $\exp\{iS\}$
over the branches of $p$.  In our approximation,
when $\eta<-\eta_{f}$ the energy is effectively time-independent and
the resulting semiclassical wave function is just the semiclassical
approximation to the product of harmonic oscillator wave functions.
In the ``frozen'' region, since $\dot{R}$, $\dot{\phi}$,
and $\dot{p_{R}}$ are $\OO(1/\sqrt{N})$, to lowest order in
our $1/\sqrt{N}$ expansion the curve $\CC_{\eta}$ will
remain frozen.  Within our approximation we may write the
classical action as:
\begin{eqnarray}
\lefteqn{S\left(R_{n}^{k},\phi_{n}^{k},\eta\right)=
\sum_{k,n}\left\{l_{n}^{k}\phi_{n}^{k} +
\int_{R_{n,0}^{k}}^{R_{n}^{k}} p_{R,n}^{k}(R_{n}^{k}, \eta_{0})\right\}}
\nonumber     \\
&&- \int_{-\infty}^{\eta}d\eta'\sqrt{\sum_{k,n} \left\{(E_{n}^{k})^{2}
+\left(nC(\eta')^{2}-n^{2}\right)(R_{n}^{k})^{2}
\right\} + C(\eta')(\xi \RR(\eta') - m_{0}^{2})}\ .\nonumber \\
\label{oscact}
\end{eqnarray}
The curvature term in the expression for $E$ is $\OO(1/N)$ smaller than the
other terms, so we shall ignore it.  As we leave the frozen
region, we may calculate the negative frequency component as
before (the reader may easily check that the turning
point is well within the region that the ``frozen approximation''
is valid), and we find once again that the Bogolubov coefficient is:
\begin{equation}
|\beta_{\{n\}}|^{2} = e^{-2\pi E_{-}/\rho}\ ,
\end{equation}
which is essentially
identical to the expression given in \pref{refasymp}.  This is
independent of the position, despite the fact that $\int d\eta E$,
and the turning point in the complex $\eta$ plane,
depends on $R$.  For more general scale factors the imaginary
part of the integral around the turning point of $E$ will depend on $R$.
In addition the time integral in the ``frozen'' regime has some
position dependence, so that the wave function begins at
$\eta = - \eta_{f}$ as a stationary state and in the
``frozen'' region begins to develop additional structure.  Nonetheless,
within the approximation we are working in, the position dependence
of ${\rm Im} \int d\eta E$ and the perturbations of the initial
wavefunction which develop as we move through the ``frozen'' region
are subleading.  One can see this by noting that
\begin{equation}
\frac{\p E}{\p R_{n}^{k}} = \frac{n^{2} \left(C(\eta)^{2}- 1\right)R_{n}^{k}}
{E}\ ;
\end{equation}
thus the change in the energy arising from the change in any given
oscillator coordinate will be
\begin{equation}
E'\delta R \sim \OO(1/\sqrt{N})\ .
\end{equation}
For computing
the integral ${\rm Im}\int d\eta E$, summing the squares of this
error gives us a total error of $\OO(1)$, compared with a lowest order answer
$E_{-}/\rho$ which is $\OO(N)$.  In examining the perturbations
to the stationary state wavefunction that accrue during the evolution
through the ``frozen'' region, we can see that for a change in a
single oscillator coordinate,
\begin{equation}
\delta \int d\eta E \sim 2 \eta_{f} \delta E \sim
\frac{|B|\ln\left\{|B|\rho\sqrt{N}\right\}}{\rho\sqrt{N}}\ ,
\end{equation}
while the integral $\int dR_{n}^{k} p_{n}^{k}$ is of $\OO(1)$
in the $1/N$ expansion, so that corrections to our description of
the wavefunction as being frozen in this regime are indeed subleading.

After the string leaves the frozen region, the curve $\CC_{\eta_{f}}$
will generally intersect some band of energy shells of the
Hamiltonian describing the evolution beyond $\eta > \eta_{f}$,
as shown in Figure 4.
The curve $\CC_{\eta>\eta_{f}}$ becomes time dependent and
starts to rotate and stretch with a periodicity given by that
of the final oscillator Hamiltonian.

We still need to check that our semiclassical approximation is valid,
by checking Equations \pref{wkbcon1} and \pref{wkbcon2}.
Hamilton-Jacobi theory tells us that $\p S/\p q = p(q)$.
Thus, the condition for Equation \pref{wkbcon1} to hold is
that:
\begin{equation}
\left|\frac{\p p_{R}}{\p R}\right| =
\left|\frac{-n^{2}C(\eta)^{2}R_{n}^{k}}{p_{n}^{k}}\right|
\ll \left(p_{n}^{k}\right)^{2}\ .
\end{equation}
This is just the usual condition for the semiclassical approximation
to hold
for the harmonic oscillator wavefunction (though in the relativistic
case this means $\omega^{2}q \ll p^{3}$ rather than
$\omega^{2} q \ll  p^{2}$); it will work for large
energies so long as we are sufficiently
far from the turning points.  The energy is large and fixed
for $\eta<\eta_{f}$ in our approximation; at $\eta=\eta_{f}$
the curve $\CC_{\eta_{f}}$ intersects a finite band of
energy shells of the final Hamiltonian, all with energy
larger than the initial energy if the scale factor expands
(and smaller if it contracts, though so long as the change in $\omega$
is not too drastic the final range of energies should still be large).
For states with oscillator modes of a given energy $E_{k,n}$,
$p_{R}$ is given by Equation \pref{pofq}, and thus Equation
\pref{wkbcon1} holds so long as $E_{k,n}$ is large.  In
our approximation this is the case; if $E_{k,n}$ was small we could treat that
oscillator via the adiabatic expansion.  To check Equation \pref{wkbcon2},
note that
\begin{equation}
\frac{\p S}{\p \eta} = E \ ,\hbox{  and  } \frac{\p^{2} S}{\p \eta^{2}}
= \frac{C_{\eta}}{C}\frac{\sum_{k,n} n^{2} C(\eta^{2}) (R_{n}^{k})^{2}}
{E}\ .
\end{equation}
Equation \pref{wkbcon2} holds if
\begin{equation}
\frac{E_{\eta}}{E^{2}} \sim \frac{\rho}{\sqrt{N}} \ll 1\ ,
\end{equation}
which is certainly the case.
One may also work with the path integral of the wave function
and check that the quadratic fluctuation operator,
given by second variation of the action along the classical
trajectory, contains terms of order $E$, so that the saddle point
integration around the classical trajectories gives a good
approximation.  Recall
also that it is the divergences in the determinant of
this quadratic fluctuation operator which contribute to
the $\beta$-function of the 2-d $\sigma$ model; we would
thus like to reiterate that although our background does
not lead to a vanishing $\beta$-function, the effects of
this lack of conformal invariance will appear in the determinant
of the quadratic fluctuation operator, while the physics we
are interested in resides in the exponential part of the
wave function, $\exp\{i S_{{\rm cl.}}\}$.

For $\eta>\eta_{f}$, the oscillator frequency has changed and the
system now has frequency $\omega_{+}$.  It is no longer in a state of
definite energy, however.  As shown in Figure 4, the curve $\CC_{\eta_{f}}$,
which is still in our approximation identical to $\CC_{-\infty}$,
intersects many tori defined by $E=constant$ when $\omega=\omega_{+}$
and so $\CC_{\eta>\eta_{f}}$ will be time dependent.
However, as we know that the
wave function is stationary for $\eta<\eta_{f}$ within
our approximation, it is easier to write
the spatial part of the wave function as the product
of Laguerre polynomials that our
semiclassical wave function approximates;
we may then simply decompose this wave function
into energy eigenstates of the Hamiltonian for $\eta>\eta_{f}$.
The
reader has doubtless noted that this is essentially the sudden approximation
for the spatial part of the wave functions.
We may argue for this approximation more directly using the Hartree
approximation for the wave functions: that is, let
\begin{equation}
\Psi = \prod_{k,n} \psi_{k,n}(R_{n}^{k},\phi_{n}^{k},\eta)\ ,
\label{hartree}
\end{equation}
and assume that correlations in the wave function are of higher
order in $1/N$.
We may write the wave equation applied to Equation \pref{hartree}
\begin{equation}
\left\{- \frac{\p^{2}}{\p\eta^{2}} +
\sum_{k,n}\HH_{k,n}^{{\rm spatial}}(R_{n}^{k},\eta)\right\} \Psi = 0
\end{equation}
as
\begin{eqnarray}
\sum_{k,n}\left\{ - \ddot{\psi}_{k,n} \prod_{(l,m)\neq (k,n)}
\psi_{l,m} \right.&-&
\dot{\psi}_{k,n}\sum_{(l,m)\neq (k,n)}\dot{\psi}_{l,m}
\prod_{(j,p)\neq (k,n),(l,m)}\psi_{j,p}    \nonumber \\
&+&\left.\HH_{k,n} \psi_{k,n} \prod_{(l,m)\neq (k,n)}
        \psi_{l,m} \right\}\ .
\end{eqnarray}
The resulting equations of motion are
\begin{equation}
\left\{-\frac{\p^{2}}{\p\eta^{2}}-N K_{k_{0},n_{0}}(\eta) \frac{\p}{\p\eta}+
\HH_{k_{0},n_{0}}(R,\eta)\right\}\psi_{k_{0},n_{0}}=0
\label{hartreewave}
\end{equation}
where
\begin{equation}
N K_{k,n}(\eta) = \sum_{(l,m)\neq(k,n)} \int \psi_{l,m}^{*} \frac{\p}{\p \eta}
        \psi_{l,m}\ .
\end{equation}
Let us rescale $\eta\rightarrow N\eta$.  The first
term is now $\OO(1/N^{2})$ and we shall drop it.
The time dependences of $K$ and of
the frequencies in $\HH$ have the form $N\eta$, so we have
a non-relativistic harmonic oscillator which varies rapidly compared
to the other time scales of the problem.
In particular,
$C(N\eta)^{2}=A+B\tanh(N\rho\eta)$ is effectively a
step function when $N\rho\gg 1$, so the sudden approximation
is valid.  This argument is admittedly
very rough but seems to capture the right physics.  A
string with many oscillators highly excited will be heavy
and slow; therefore, it will be nonrelativistic and any change
in the spacetime, if moderate, will occur on a time scale much smaller than
the response time of the string.  We should note that we still need the
relativistic analysis above to find the string production amplitudes,
as there is no string production in the nonrelativistic approximation.

\figloc4{Evolution of one of the Fourier modes of the string.
For $\eta < \eta_{f}$  $p_{k}(q_{k})$ is defined by the
solid curve $\CC_{-\infty}$,
which is a curve of constant energy $E^{k}_{-}$ in phase space.
For $\eta < \eta_{f}$ the curves of constant energy are the
dashed ellipses.  The dotted curve $\CC_{\eta>\eta_{f}}$
denotes the function $p(q)$ found by evolving $\CC_{-\infty}$
from $\eta_{f}$ to some $\eta$ using the string Hamiltonian.}

We have thus shown that in this model, the number of strings per unit volume
produced in a given state with energy $E_{-}$ is proportional to
$\exp\{-2\pi E_{-}/\rho\}$.  At sufficiently large oscillator level
$N$, if the center of mass momentum is of order 1,
$E_{-}\approx m\approx\sqrt{N/\apr}$ where $m$ is the mass.
It is well known that the
number of string states of mass $m$, for $m\gg1/\apr$,
can be approximated by the
formula
\begin{equation}
D(m) \approx m^{-b} e^{c\sqrt{\apr}m}\ ,
\label{dofstates}
\end{equation}
where $b$ and $c$ are pure positive numbers of order 10
and depend on the specific string theory one is using.  So far we have
calculated the production amplitude for an arbitrary state of a
given mass regardless of whether or not it was a physical state.  The
density of states should automatically count physical states,
so if we use it to
count the total number of strings produced we should
get the correct answer.  Thus the total energy density
produced during the expansion
is, if we assume that our production formula holds for most
string states at a given mass,
\begin{equation}
\frac{\dot{\EE}_{{\rm total}}}{{\rm unit volume}}
\approx \int_{0}^{\infty}dm\ mD(m)e^{-2\pi m/\rho}\ .
\label{edens}
\end{equation}
If we allow ourselves the luxury of extrapolating our result
to expansion rates of order the inverse string scale $\sqrt{\apr}$,
we find that when
\begin{equation}
\rho > \frac{2\pi}{c\sqrt{\apr}}
\end{equation}
the integral in \pref{edens} blows up in a manner reminiscent of the
Hagedorn transition.  This is the main result of this paper.

\subsection{Interpretation}

The fact that the energy produced at a critical value of the
expansion rate diverges indicates to us that computing string production
on a fixed background leads to a diverging rate of string production.
We believe this divergence to be a sign that we
have either reached some sort of phase transition in the theory,
or we have neglected the effects of the backreaction of
the produced string matter.
We may speculate in the latter
case that since the bulk of the energy is produced
in very large strings as $\rho$ approaches $1/\sqrt{\apr}$, the
backreaction should act to slow the expansion, as we have previously
claimed should occur with a sufficient density of strings which
grow with the scale factor.  Since this problem will arise whenever
the expansion rate is sufficiently fast, we argue that the
effect of backreaction will be to sufficiently slow the expansion
so that any curvature singularity, if reached, is reached at infinite proper
time and that all geodesics become
completeable.

\section{Speculations about anisotropically expanding models}

The arguments in the previous section are far from a rigorous solution
to the problem of generic singularities, which as we have
argued are very anisotropic and furthermore have no ``out''
region with an unambiguous definition of particle number.  As
for the latter, we hope that the general features of
our answer in terms of the more covariant
concept of the energy-momentum tensor will carry through to
a calculation of this quantity in a background with
diverging rates of expansion and contraction, even though
the details of the calculation will be very different; in particular the
essential singularity in the calculated
rate of energy production seems physically well-motivated and general,
being based on the fight between
an exponential suppression of string production
and the exponential growth of the density of states.

If we wish to understand the further evolution
of the spacetime, we need to understand how
the anisotropic expansion and contraction
which exists near generic singularities
affects the configuration of produced strings, and
we need to understand the equation of state in these
anisotropic spacetimes.  For example,
one might expect that strings are preferentially produced
with excitations in the most rapidly expanding and
contracting directions, and that most of the strings produced will in
the expanding case be large strings growing as the
scale factor and in the contracting case be small strings with
large momentum in the oscillator modes.  The wave functions
we have constructed are stationary states with no definite shape.
To make our scenario plausible we would need to work with some
set of coherent states which describe strings in definite classical
configurations; since the highly massive strings are
generally nonrelativistic the width of the string wave function
around the peak configuration will not spread too quickly.
We have not been able to detect any preference in the production
of any particular configuration at our level of approximation,
even in the isotropically expanding case.

As for the question of which string field modes are preferentially produced,
if we examine strings which are polarized in a direction
or set of directions with a given expansion rate, the
number of strings produced will increase with the
expansion rate in that direction.  On the other hand,
strings with no polarization in the expanding and contracting directions
(if some spatial directions have a constant scale factor)
should not be produced at all.  There should be some
extrapolation between these extremes, but our approximations
seem to be relatively insensitive to the distribution of modes
among the various polarizations.
Clearly there
is a considerable amount of work to do.

One natural scenario for the evolution of the spacetime is
the following: as we
approach the singularity, where it is approximated by the
Kasner metric, string production becomes important,
and we produce unstable strings preferentially polarized in the
directions which are changing the most rapidly.  The
equation of state for the growing strings is incompatible
with a constant dilaton, so the equations governing
string cosmology, Equations \pref{stcoseq}, effectively describe the
spacetime.  As $\dot{\lambda}_{k}$ approaches
the string scale, the energy becomes extremely large;
the expanding directions feel a large negative pressure and
the contracting directions feel a large positive pressure,
and the spacetime relaxes to a non-singular evolution.

What in fact happens will depend on the correct expressions for
$\rho$ and $p$, the initial value of $\phi$, and higher order
corrections to the $\beta$-function equations.  Based on some
simple computer simulations and qualitative arguments, we have
found several possible outcomes for evolution via Equations
\pref{stcoseq}.  There seems to be little possibility for the
rapid inflation and contraction to turn around before the expansion rate
reaches unity in string units.  What happens when
it reaches this point depends
on how we interpret the diverging integral in Equation \pref{edens}.
If we permit ourselves to imagine that the effect is a sufficiently large
production of string matter then the expansion and contraction will
suddenly reverse due to the large pressures we expect to arise
from strings in rapidly expanding and contracting spacetimes.
Once the turnaround occurs much will depend on the equation of state for
less rapidly changing spacetimes.  If expansion halts completely
and the pressure drops to zero, then the spacetime will stabilize
at a constant scale factor but the large energy density produced will
cause the dilaton to grow to infinity in finite time, so that string loop
effects become extremely important.  We might imagine using
the conjectured duality properties of string theory
(see for example Sen 1994, Hull and Townsend 1995, Duff 1995, Witten 1995)
at this point to relate the background to the weakly coupled
background of another theory.  Another scenario occurs if
the string coupling is
too weak when the expansion rate becomes of order unity and
not enough energy is produced; the large continued contraction
of the angular directions can cause
the otherwise growing dilaton to return to weak coupling as we reach the
singularity.  Of course, in the regime we are discussing all
quantities -- $\rho$, $p_{k}$, $\dot{\lambda}_{k}$, and so
on -- are of order one in string units; we can only
speculate what the form of the $\beta$-function equations
and the equation of state might be. In discussing the form of $\rho$ and
$p$ we should note also that at the point that the expansion becomes
of order unity, the singularity in the Schwarzschild geometry is reached
within a single string time.  It is not clear that
strings can equilibrate.  Nonetheless,
the blow-up of the integral in Equation \pref{edens} indicates
that the description of the region near the putative singularity
may be drastically modified by the quantum dynamics of
string field theory.

\section{Conclusions}

We have argued that in regions of rapid spatial expansion, such as the
region close to
a spacelike singularity,
a divergent energy density of string is produced as the curvature
approaches the string scale.  This leads us to one of two
conclusions, both of which are interesting and merit further
investigation:  either string theory the region close to the singularity
undergoes a phase transition as the density of string approaches
one string mass per string volume, or
properly taking into account
the back reaction of the produced strings on the spacetime
pushes the singularity to infinite proper time.
The answer could also be some combination of the two, or it could be that
string theory is just inconsistent, although we feel that the
divergence calculated has the flavor of a divergence
arising from an incorrect approximation or
extrapolation (\ie\ from neglecting either backreaction
or a phase transition at large densities), and so
there is no obvious reason to call this divergence a symptom of some
disease of the underlying theory.

If the backreaction becomes important in the manner we have suggested, it
is an indication that we must take seriously unstable macroscopic
strings as sources in the $\beta$-function equations.  This
not to say that solving the $\beta$-function equations gives
us an inconsistent picture, but rather that high order loop effects
become important in spacetimes with unstable strings.
It would be interesting to find some sort
of self-consistent way to include
macroscopic strings as sources directly in the $\beta$-function
equation, so that one could evolve the spacetime accordingly
and see what happens to singular regions and to the singularity theorems.
One might also envision trying to construct this approximation directly
by resumming certain classes of loop corrections to the $\beta$-function
equations.

Understanding how string theory modifies the dynamical equations of
spacetime is extremely important for a resolution of the
problem of black hole evaporation.  A transition to a uniformly expanding
region (Frolov \etal\ 1990); continued nonsingular contraction of the
two-spheres and
expansion of the radial direction
(Martinec 1995) (where ``radial'' refers to the spacelike direction in the
interior of the black hole);
or simultaneous collapse of the expanding directions
are all possible depending on the density of state and the
equations of motion of the background.  The first two possibilities
tell us that the information is carried away to inaccessible regions
of spacetime; the third possibility implies a slowly-evaporating-remnant
scenario, where information is brought near the horizon by the recollapse
of the radial scale factor.

String production in a non-trivial background spacetime should be an
approximation to the full quantum dynamics of the vacuum condensate.
It may be that we have shown that it is incorrect not only
in principle but in practice to separate dynamical quantum string fields
and classical string backgrounds.  If this is the case then we are
stuck until we develop a manifestly background independent understanding
of string field theory, or its nonperturbative completion.

\vskip .2in
{\bf Acknowledgements}

We would like to thank Eanna Flanagan, Vivek Iyer, and Martin
O'Loughlin for enlightening conversations and helpful suggestions.
A.L. and E.M. are supported in part by funds provided by the DOE
under grant No. DOE-FG02-90ER-40560.

\appendix
\section{Matching 4d black holes to 2d black holes}

Although we have argued that classical (tree level) string theory
is insufficient for understanding spacelike singularities, it would be
nice to better understand the classical geometry of singularities
in string theory,
in order to have some confidence in our starting point and
to sharpen some of the arguments in the Introduction regarding
the existence of singularities in string theory.
We would like to repeat and extend some arguments
made by Giddings \etal\ (1993, 1994) concerning the
worldsheet renormalization group flow of the $\sigma$-model,
here without the benefit of any gauge field or axion hair to
stabilize the shrinking two-spheres.

Close to the singularity the spacetime metric should look like
\begin{equation}
        ds^{2} = -dT^{2} + a^{2}(T) d\rho^{2} + r^{2}(T)d\Omega^{2}\ .
\end{equation}
In the classical geometry we approach the singularity in
finite time as the metric in the $\rho$ direction blows
up,
and we must go beyond the lowest order solution to the
$\beta$-function equations.  It is well-known that the
$S^{2}$ part of the $\sigma$-model should flow to a
trivial $c=0$ theory.  Cecotti and Vafa (1992) have
computed the Zamolodchikov metric along the entire
renormalization group flow of the supersymmetric $CP^{1}=S^{2}$
model
and found the metric to be well defined, even as $r^{2}$
becomes negative (see also Witten 1993).
As the radius gets small and passes
through zero the theory masses up and becomes trivial,
with the soliton masses increasing as $m = 8\exp (-r^{2}/2)$.
Once the radius is sufficiently negative, the
$T,\rho$ plane will have excess central charge; a natural
fixed point of the renormalization group flow is
the $SL(2,\RR)/U(1)$ black hole.  This would mean that the
velocity of the transverse stretching direction dominates
the $\beta$-function for $a$.  One might convince oneself
of this by adding a potential term to the spacetime action
simulating the effects of string instantons wrapping around the $S^{2}$
direction (as do Giddings \etal\ (1994)).  This potential should
be of the form
\begin{equation}
V(R,\phi) = e^{-2\phi - 2 r^{2}/\apr}
\end{equation}
(Giddings \etal\ 1994) where $\phi$ is the dilaton.  This is a function
of the scale factor which is of order unity at zero radius: it
cannot compete with the large $\dot{\lambda}$ terms present
near a velocity-dominated singularity.

This is an attractive scenario, but it has not been proven.  Exact
results such as those of Cecotti and Vafa (1992) do not help us much.
The contribution of the $S^{2}$ degrees of freedom to the
dilaton $\beta$-function should be given by the Zamolodchikov
$c$-function of the $S^{2}$ theory (Fradkin and Tseytlin 1985,
Callan \etal\ 1985, Polyakov 1986, 1987), which is only simply related
to the Zamolodchikov metric at RG fixed points (Zamolodchikov 1986).
Even if we had this information, cross-couplings between $a$ and
$r$ would make the calculation exceedingly difficult.

We can also try to work backwards from the singularity, starting with the
WZW black hole at small small $r$, (where $r$ is the
timelike direction inside the horizon, and $t$ is the spacelike coordinate),
times the $S^{2}$ theory close to its trivial infrared fixed point;
with these initial conditions, the velocity terms coming from the
coupling of the $S^{2}$ are small and we have some intuition for
the evolution of the $\beta$-function equations.
Here we would use the following procedure.  Begin
with a spacelike slice $\Sigma$ in this region.  Given
the mass of the WZW black hole, a starting point inside the horizon,
and a starting point in the RG trajectory of the $S^{2}$ theory, we
have a set of initial conditions for evolving the
$\beta$-function equations of the $\sigma$-model away from the singularity in
the radial direction.  We may do this, for example, by using the
well-known analogy between the $\beta$-function equations in
spacetime and the RG flow of the spatial part of the $\sigma$-model,
where in the latter picture we would gravitationally dress the
spatial $\sigma$-model with appropriate functions of the Liouville
mode (Polchinski 1989,
Das \etal\ 1989, Banks and Lykken 1990).  In the beginning
of the radial evolution we find that the spacetime is some small perturbation
of the WZW black hole coupled to the $S^{2}$ theory moving
away from the trivial fixed point.  If we begin with the $S^{2}$
theory close enough to this fixed point,
as we pass through the horizon at $r=r_{h}$ the geometry will still
be described by the WZW black hole times the massive $S^{2}$ theory
with the mass slowly decreasing, and the $r-t$ theory
approaches the asymptotically flat linear dilaton vacuum.
At some point $r=r_{c}$ the $S^{2}$ theory
will begin crossing over to behavior characterized by the ultraviolet,
asymptotically free fixed point and we should enter an asymptotically flat
regime with a constant dilaton (since there is no longer any
excess central charge to feed a linear dilaton solution).  At this point,
by Birkhoff's theorem, any deviations from the Minkowski metric should be
that arising from the Schwarzschild solution at large radius.  If we evolve
the geometry from $\Sigma$ towards the singularity, as long as
the $S^{2}$ theory sufficiently close to its IR fixed point,
we expect to reach the singularity of the WZW black hole, in finite time,
since the effect of the flow of the $S^{2}$ theory should be minimal.
The geometry
is then that of a long throat with an effectively 2-dimensional geometry,
terminating at one end at a 2d black hole singularity and opening out at the
radius $r=r_{c}$, well outside the horizon of the 2d theory, into an
asymptotically flat 4-dimensional spacetime.  It may be that when we
reach $r_{c}$ the crossover forces us into some other fixed point,
but we believe this to be unlikely.

Now let us move in the space of $c=4$ conformal field theories by
moving the initial position of the $S^{2}$ theory farther away from
its trivial fixed point.  This should have the effect of increasing
the initial velocity of the coupling of the $S^{2}$ theory to $r$,
which making cross-coupling
terms between the radial and angular directions more important.
We then expect $r_{c}$ to approach $r_{h}$; eventually along
this marginal line, $r_{h}$ enters the asymptotically
flat regime, unless we are driven by the crossover behavior
to some other fixed point.  Since at this point, the velocity terms coming from
$S^{2}$ at $\Sigma$ should be fairly large, we no longer expect
the region near the singularity to look like the WZW black hole;
the RG flow might push the WZW singularity off to infinite
distance, but we expect instead that we have some standard velocity-dominated
singularity, based on our previous arguments.

\vskip 1cm
\leftline{\bf \large \noindent References}
\vskip .5cm
\begin{itemize}
\itemindent=-18pt\itemsep=0pt\parsep=0pt

\item[] Amati D and Klimcik C 1989 \pl\ B {\bf 219} 443
\item[] Aspinwall P S, Greene B R, and Morrison D R 1993
\pl\ B {\bf 303} 249
\item[] Aspinwall P S, Greene B R, and Morrison D R 1994
\np\ B {\bf 416} 414
\item[] Audretsch J 1979 \jp\ A {\bf 12} 1189
\item[] Banks T 1994 {\it Lectures on Black Holes and Information
Loss (Lectures from the Spring School on Supersymmetry, Supergravity and
Superstrings (Trieste))}\ hep-th/9412131
\item[] Banks T and Lykken J 1990 \np\ B {\bf 331} 173
\item[] Belinskii V A, Khalatnikov I M, and Lifshitz E M 1970
\adv\ {\bf 19} 525
\item[] Belinskii V A, Khalatnikov I M, and Lifshitz E M 1982
\adv\ {\bf 31} 639
\item[] Bernard C and Duncan A 1977 \ap\ {\bf 107} 201
\item[] Berry M V 1981 {\it Chaotic Behavior of Deterministic Systems},
Les Houches summer school proceedings (Amsterdam, North-Holland)
\item[] Berry M V and Balazs N 1979 \jp\ A {\bf 12} 625
\item[] Birrell N D and Davies P C W 1982  {\it Quantum Fields
in Curved Space} (Cambridge, UK: Cambridge University Press)
\item[] Callan C G, Friedan D, Martinec E J, and Perry M P 1985
\np\ B {\bf 262}
\item[] Callan C G, Giddings S B, Harvey J A, and Strominger A 1992
\pr\ D {\bf 45} R1005
\item[] Callan C G and Khuri R R 1991 \pl\ B {\bf 261} 363
\item[] Callan C G, Lovelace C, Nappi C R, and Yost S A
1987 \np\ B {\bf 288} 525
\item[] Cecotti S and Vafa C 1992 \prl\ {\bf 68} 903
\item[] Dabholkar A and Harvey J A 1989 \prl\ {\bf 63} 719
\item[] Dabholkar A, Gibbons G, Harvey J A, and Ruiz-Ruiz F 1990
\np\ B {\bf 340} 33
\item[] Das S, Naik S, and Wadia S 1989 \mpl\ A {\bf 4} 1033
\item[] de Vega H and Sanchez N 1992 \pr\ D {\bf 45} 2783
\item[] DeWitt B S 1957 \rmp\ {\bf 29} 377
\item[] Dijkgraaf R, Verlinde H, and Verlinde E 1992 \np\ B {\bf 371} 269
\item[] Dixon L J, Harvey J A, Vafa C, and
Witten E 1985 \np\ B {\bf 261} 678
\item[] Dixon L J, Harvey J A, Vafa C, and
Witten E 1986 \np\ B {\bf 274} 285
\item[] Duff M 1995 \np\ B {\bf 442} 47
\item[] Eckhardt C 1930 \pr\ {\bf 35} 1303
\item[] Epstein H, Gaser V, and Jaffe A 1965 \nc\ {\bf 36} 1016
\item[] Fischler W and Susskind L 1986 \pl\ B {\bf 171} 383,
\pl\ B {\bf 173} 262
\item[] Fradkin E S and Tseytlin A A 1985 \np\ B {\bf 261} 1
\item[] Friedan D 1985 \pl\ B {\bf 162} 102
\item[] Frolov V P, Markov M A, and Mukhanov V F 1990
\pr\ D {\bf 41} 383
\item[] Fulling S A 1989 {\it Aspects of Quantum Field Theory in
Curved Spacetimes} (Cambridge, UK: Cambridge University Press)
\item[] Gasperini M, Sanchez N, and Veneziano G 1991a
\ijmp\ A {\bf 6} 3853
\item[] Gasperini M, Sanchez N, and Veneziano G 1991b
\np\ B {\bf 364} 365
\item[] Gauntlett J P, Harvey J A, Robinson M M, and
Waldram D 1994 \np\ B {\bf 411} 461
\item[] Giddings S B, Polchinski J, and Strominger A 1993
\pr\ D {\bf 48} 5784
\item[] Giddings S B, Harvey J A, Polchinski J G, Shenker S H,
and Strominger A 1994 \pr\ D {\bf 50} 6422
\item[] Gross D J and Mende P F 1987 \pl\ B {\bf 197} 129
\item[] Gross D J and Mende P F 1988 \np\ B {\bf 303} 407
\item[] Hawking S W 1967 \prcl\ A {\bf 300} 182
\item[] Hawking S W 1974 {\it Nature} (London) {\bf 248} 30
\item[] Hawking S W and Penrose R 1970 \prcl\ A {\bf 314} 529
\item[] Horowitz G T and Myers R 1995 {\it The value of singularities}\ UCSB
preprint UCSBTH-95-6, gr-qc/9503062
\item[] Horowitz G T and Steif A R 1990, \prl\ {\bf 64} 260,
\pr\ D {\bf 42} 1950
\item[] Hull C M and Townsend P K 1995 \np\ B {\bf 438} 109
\item[] Johnson C V 1994 \pr\ D {\bf 50} 4032
\item[] Khuri R R 1993 \pl\ B {\bf 307} 298, \np\ B {\bf 403} 335,
\pr\ D {\bf 48} 2823
\item[] Landau, L D and Lifshitz E M 1977, {\it
Quantum Mechanics: non-relativistic theory, 3rd. ed.} (NY: Pergamon)
\item[] Lifshitz E M and Khalatnikov I M 1963 \adv\ {\bf 12} 185
\item[] Lovelace C 1986 \np\ B {\bf 273} 413
\item[] Lowe D and Strominger A 1994 \prl\ {\bf 73} 1468
\item[] Martinec E J 1995 \cqg\ {\bf 12} 941
\item[] Maslov V P and Fedoriuk M V 1965, {\it Semiclassical Approximation
in Quantum Mechanics} (Dordrecht: Reidel) (original Russian edition 1965)
\item[] Migdal A B 1977 {\it Qualitative Methods in Quantum Theory}
(Reading MA: W.A. Benjamin)
\item[] Nappi C and Witten E 1992 \pl\ B {\bf 293} 309
\item[] Penrose R 1965 \prl\ {\bf 14} 57
\item[] Peremolov A M and Popov V S 1969 \jetp\ {\bf 39} 738
\item[] Poisson E and Israel W 1988 \cqg\ {\bf 5} L201
\item[] Polchinski J 1989 \np\ B {\bf 324} 123
\item[] Polyakov A M 1986 {\it Phys. Scripta}\ T{\bf 15} 191
\item[] Polyakov A M 1987 {\it Gauge Fields and Strings}
(New York: Harwood Academic)
\item[] Pokrovskii V I and Khalatnikov I M 1961
\jetp\ {\bf 13} 1207
\item[] Sanchez N and Veneziano G 1990 \np\ B {\bf 333} 253
\item[] Sauter 1932 {\it Z Phys.}\ {\bf 73} 547
\item[] Schoen T and Yau S-T 1983 \cmp\ {\bf 90} 575
\item[] Sen A 1985 \pr\ D {\bf 32} 2102, \prl\ {\bf 55} 1846
\item[] Sen A 1994 \ijmp\ A {\b 9} 3707
\item[] Strominger A 1995 {\it Les Houches Lectures on Black Holes},
Lectures the Les Houches Summer School,
``Fluctuating geometries in statistical mechanics and
field theory'', Les Houches, France,
hep-th/9501071
\item[] Tseytlin A A 1990 \pl\ B {\bf 251} 530
\item[] Tseytlin A A 1992 \cqg\ {\bf 9} 979
\item[] Tseytlin A A and Vafa C 1992 \np\ B {\bf 372} 443
\item[] Veneziano G 1991 \pl\ B {\bf 265} 287
\item[] Witten E 1991 \pr\ D {\bf 44} 314
\item[] Witten E 1993 \np\ B {\bf 403} 159
\item[] Witten E 1995 \np\ B {\bf 443} 85
\item[] Zamolodchikov A B 1986 \jetpl\ {\bf 43} 730
\end{itemize}

\end{document}